\documentclass[times, twocolumn, twocolappendix]{aastex631}

\usepackage{savesym}
\savesymbol{tablenum}
\usepackage{siunitx}
\restoresymbol{SIX}{tablenum}

\usepackage{tipa}
\usepackage{amsmath}
\usepackage{ulem}

\begin{document}

\title{Collisional growth efficiency of dust aggregates and its independence of the strength of interparticle rolling friction}

\author[0000-0003-0947-9962]{Sota Arakawa}
\affiliation{Japan Agency for Marine-Earth Science and Technology, 3173-25, Showa-machi, Kanazawa-ku, Yokohama, 236-0001, Japan}

\author[0000-0001-9659-658X]{Hidekazu Tanaka}
\affiliation{Astronomical Institute, Graduate School of Science, Tohoku University, 6-3 Aramaki, Aoba-ku, Sendai 980-8578, Japan}

\author[0000-0002-5486-7828]{Eiichiro Kokubo}
\affiliation{National Astronomical Observatory of Japan, 2-21-1, Osawa, Mitaka, Tokyo, 181-8588, Japan}



\begin{abstract}
The pairwise collisional growth of dust aggregates consisting submicron-sized grains is the first step of the planet formation, and understanding the collisional behavior of dust aggregates is therefore essential.
It is known that the main energy dissipation mechanisms are the tangential frictions between particles in contact, namely, rolling, sliding, and twisting.
However, there is a large uncertainty for the strength of rolling friction, and the dependence of the collisional growth condition on the strength of rolling friction was poorly understood.
Here we performed numerical simulations of collisions between two equal-mass porous aggregates with various collision velocities and impact parameters, and we also changed the strength of rolling friction systematically.
We found that the threshold of the collision velocity for the fragmentation of dust aggregates is nearly independent of the strength of rolling friction.
This is because the total amount of the energy dissipation by the tangential frictions is nearly constant even though the strength of rolling friction is varied.
\end{abstract}



\section{Introduction}

The first step of the planet formation is the pairwise collisional growth of dust aggregates consisting submicron-sized grains \citep[e.g.,][]{2021NatRP...3..405W}.
Therefore, understanding the condition for collisional growth of dust aggregates is essential.

The collisional behavior of dust aggregates has been studied extensively, by using both laboratory experiments \citep[e.g.,][]{2008ARA&A..46...21B, 2021ApJ...923..134F, 2022MNRAS.509.5641S} and numerical simulations \citep[e.g.,][]{2009ApJ...702.1490W, 2021ApJ...915...22H, 2021A&A...652A..40U}.
It is known that the threshold collision velocity for the fragmentation of dust aggregates, $v_{\rm fra}$, depends on the strength of interparticle forces acting on constituent particles in contact.

The stickiness of dust particles has been investigated in a large number of studies \citep[e.g.,][]{2012arXiv1204.0001G, 2012PThPS.195..101T, 2015ApJ...798...34G}.
These studies found that the threshold velocity for the sticking in a head-on collision of individual dust particles is several times higher than that predicted for perfectly elastic spheres.
\citet{2021ApJ...910..130A} concluded that the viscous dissipation for the normal motion would play a critical role in collision between micron-sized water ice particles.

Recently, \citet{2022ApJ...933..144A} performed numerical simulations of collisions between two equal-mass dust aggregates consisting submicron-sized spherical ice particles.
Surprisingly, they found that the main energy dissipation mechanism is not the viscous dissipation but the friction between particles in contact.
Interparticle frictions are caused by tangential motions, and there are three types of tangential motions: rolling, sliding, and twisting \citep[see Figure 2 of][]{2007ApJ...661..320W}.
Therefore, we can imagine that the collisional behavior of dust aggregates would be affected by the strength of interparticle frictions.

The rolling friction is one of the strongest energy dissipation mechanisms.
The interparticle rolling motion behave elastically (i.e., without energy dissipation) when the length of the rolling displacement, $\xi$, is smaller than the critical rolling displacement, $\xi_{\rm crit}$ \citep[e.g.,][]{1995PMagA..72..783D, 2007ApJ...661..320W}.
The energy dissipation starts when $\xi$ exceeds $\xi_{\rm crit}$.

As the interparticle rolling friction is the dominant mechanism for the energy dissipation within colliding dust aggregates, we can expect that the collisional outcome of dust aggregates must depend on $\xi_{\rm crit}$; however, it is poorly understood that how large $\xi_{\rm crit}$ is for micron-sized spheres. 
\citet{1995PMagA..72..783D} constructed the theoretical model of rolling based on the contact theory developed by \citet{1971RSPSA.324..301J}.
They mentioned that $\xi_{\rm crit}$ should be of the order of the distance between atoms in the materials, and it is typically $0.2~\si{nm}$.
In contrast, \citet{1999PhRvL..83.3328H} measured the rolling friction torques between individual silica microspheres by using an atomic force microscope.
They found, however, that the critical rolling displacement for silica microspheres with radii between $0.5$ and $2.5~\si{\micro m}$ would be $\xi_{\rm crit} = 3.2~\si{nm}$, which is 16 times larger than that \citet{1995PMagA..72..783D} assumed.
We note that the upper limit of $\xi_{\rm crit}$ would be given by the contact radius of spheres at the equilibrium state, $a_{0}$, and $\xi_{\rm crit} = 3.2~\si{nm}$ is still smaller than $a_{0}$.

\citet{2014JPhD...47q5302K} modified the analytical theory for the rolling friction based on the concept of adhesion hysteresis, and they reported that $\xi_{\rm crit}$ would be given by $\xi_{\rm crit} \sim a_{0} / 24$ for silica microspheres when the crack propagation rate is $1$--$10~\si{\micro m.s^{-1}}$.
Recently, \citet{2021NatSR..1114591U} performed molecular dynamics simulations of amorphous Lennard--Jones grains and confirmed that $\xi_{\rm crit}$ depends on both material parameters and angular velocity for rolling.

Although the rolling friction is one of the main energy dissipation mechanisms for oblique collisions between dust aggregates, dependence of the collisional growth condition on $\xi_{\rm crit}$ is poorly understood.
\citet{2008ApJ...677.1296W} performed numerical simulations of head-on collisions between dust aggregates with various $\xi_{\rm crit}$.
They used highly porous dust aggregates prepared by ballistic cluster--cluster aggregation in their simulations, and they investigated the condition for collisional compression and fragmentation.
They reported that both the onset of collisional compression and the maximum compression linearly depend on $\xi_{\rm crit}$, but the size of the largest fragment formed after a collision is nearly independent of $\xi_{\rm crit}$.
These results would be naturally explained by the findings that the collisonal compression is caused by interparticle rolling motions while the energy dissipation is primarily caused by connection and disconnection of particles \citep{2009ApJ...702.1490W}.
We note, however, that they focused on head-on collisions and did not study collisonal behavior for oblique collisions.
\citet{2008ApJ...684.1310S, 2012ApJ...753..115S} also investigated the collisional compression of dust aggregates by sequential collision simulations, but they also focused on head-on collisions.

In this study, we report the results from numerical simulations of collisions between two equal-mass dust aggregates with various collision velocity and impact parameters.
We changed the strength of interparticle rolling friction systematically, which is controlled by $\xi_{\rm crit}$.
Surprisingly, we found that the threshold collision velocity for the fragmentation of dust aggregates is nearly independent of $\xi_{\rm crit}$ when we consider oblique collisions.

\section{Model}

We performed three-dimensional numerical simulations of collisions between two equal-mass dust aggregates.
The numerical code used in this study is identical to that developed by \citet{2022ApJ...933..144A}, which considered the viscous drag for normal motion and also considered friction torques for tangential motions.

\subsection{Material parameters}

We assumed that dust particles constituting dust aggregates are made of water ice, and the all particles have the same radius of $r_{1} = 0.1~\si{\micro m}$.
The material parameters of water ice particles used in this study are listed in Table \ref{table1}.
The particle interaction model is identical to that used in \citet{2022ApJ...933..144A}.

\begin{table}
\caption{
List of material parameters used in this study \citep[see][]{2022ApJ...933..144A}.
}
\label{table1}
\centering
\begin{tabular}{ccc}
{\bf Parameter}                  & {\bf Symbol}       & {\bf Value}                   \\ \hline
Particle radius                  & $r_{1}$            & $0.1~\si{\micro m}$           \\
Material density                 & $\rho$             & $1000~\si{kg.m^{-3}}$         \\
Surface energy                   & $\gamma$           & $100~\si{mJ.m^{-2}}$          \\
Young's modulus                  & $\mathcal{E}$      & $7~\si{GPa}$                  \\
Poisson's ratio                  & $\nu$              & $0.25$                        \\
Viscoelastic timescale           & $T_{\rm vis}$      & $6~\si{ps}$                   \\ \hline
\end{tabular}
\end{table}

\subsection{Interparticle rolling motion}
\label{sec.roll}

In this study, we regarded $\xi_{\rm crit}$ as a parameter.
We applied a linear spring model with a critical displacement to interparticle rolling motion \citep{2007ApJ...661..320W}.
When the length of the rolling displacement is $\xi$, the potential energy stored by the rolling displacement is given by
\begin{equation}
U_{\rm r} = \frac{1}{2} k_{\rm r} \xi^{2},
\end{equation}
and the spring constant, $k_{\rm r}$, is given by
\begin{equation}
k_{\rm r} = \frac{4 F_{\rm c}}{R}.
\end{equation}
Here $F_{\rm c} = 3 \pi \gamma R$ is the maximum force needed to separate the two particles in contact, and $R = r_{1} / 2$ is the reduced particle radius.
The critical energy required to start rolling, $E_{\rm r, crit}$, is given by 
\begin{eqnarray}
\label{eq.crit-roll}
E_{\rm r, crit} & = & \frac{1}{2} k_{\rm r} {\xi_{\rm crit}}^{2} \nonumber \\
                & = & 6 \pi \gamma {\xi_{\rm crit}}^{2}.
\end{eqnarray}
\citet{2007ApJ...661..320W} introduced $E_{\rm roll}$ as the energy needed to rotate a particle by $\pi / 2$ radian around its contact point, which is given by
\begin{eqnarray}
\label{eq.roll}
E_{\rm roll} & = & k_{\rm r} \xi_{\rm crit} \pi R \nonumber \\
             & = & 12 \pi^{2} \gamma R \xi_{\rm crit}.
\end{eqnarray}

\begin{figure}
\centering
\includegraphics[width=\columnwidth]{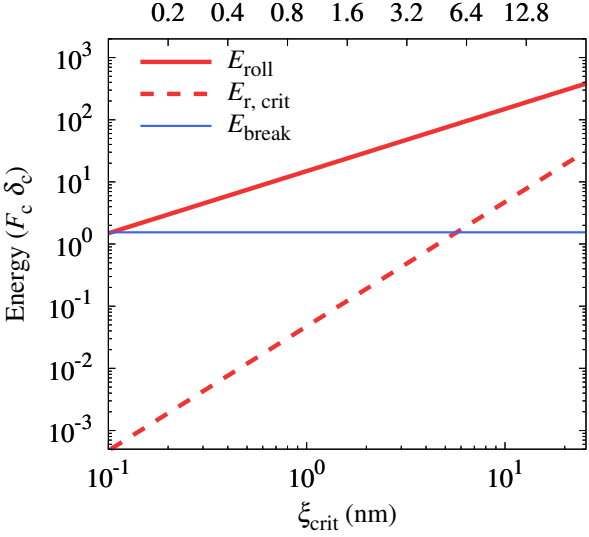}
\caption{
Dependence of $E_{\rm r, crit}$ and $E_{\rm roll}$ on $\xi_{\rm crit}$.
}
\label{fig.eroll}
\end{figure}

Figure \ref{fig.eroll} shows the dependence of $E_{\rm r, crit}$ and $E_{\rm roll}$ on $\xi_{\rm crit}$.
We normalized $E_{\rm r, crit}$ and $E_{\rm roll}$ by $F_{\rm c} \delta_{\rm c}$, where $\delta_{\rm c}$ is the critical stretching length for particle separation, which is given by
\begin{equation}
\delta_{\rm c} = {\left( \frac{9}{16} \right)}^{1/3} \frac{{a_{0}}^{2}}{3 R}.
\end{equation}
The contact radius of spheres at the equilibrium state, $a_{0}$, is given by
\begin{eqnarray}
a_{0} & = & {\left( \frac{9 \pi \gamma R^{2}}{\mathcal{E}^{*}} \right)}^{1/3} \nonumber \\
      & = & 12.4~\si{nm},
\end{eqnarray}
where $\mathcal{E}^{*} = \mathcal{E} / {[ 2 {( 1 - \nu^2 )} ]}$ is the reduced Young's modulus.

The energy needed to break a contact in the equilibrium by quasistatic process, $E_{\rm break}$, is given by
\begin{eqnarray}
\label{eq.break}
E_{\rm break} & = & {\left( \frac{4}{45} + \frac{4}{5} \times 6^{1/3} \right)} F_{\rm c} \delta_{\rm c} \nonumber \\
              & = & 1.54 F_{\rm c} \delta_{\rm c}.
\end{eqnarray}
The blue horizontal line shown in Figure \ref{fig.eroll} represents $E_{\rm break}$.
We found that $E_{\rm break} > E_{\rm r, crit}$ when $\xi_{\rm crit} \le 3.2~\si{nm}$.
Not only the potential energy related to the normal compression but also potential energies stored by tangential displacements dissipate when two particles separate.
However, the contribution of the rolling displacement is negligible for $\xi_{\rm crit} \le 3.2~\si{nm}$.
In addition, \citet{2022ApJ...933..144A} confirmed that contributions of sliding and twisting displacements also play a minor role; they are an order of magnitude smaller than $E_{\rm break}$.

The collision velocity of two dust aggregates is $v_{\rm col}$, and the initial kinetic energy per one particle with the velocity of $v_{\rm col} / 2$ is
\begin{eqnarray}
\label{eq.E1}
E_{\rm kin, 1} & = & \frac{1}{2} m_{1} {\left( \frac{v_{\rm col}}{2} \right)}^{2} \nonumber \\
               & = & 33 {\left( \frac{v_{\rm col}}{50~\si{m.s^{-1}}} \right)}^{2} F_{\rm c} \delta_{\rm c},
\end{eqnarray}
where $m_{1} = 4 \pi \rho {r_{1}}^{3} / 3$ is the mass of each particle.
Therefore, $E_{\rm roll} \gg E_{\rm kin, 1}$ for $\xi_{\rm crit} \gg 3.2~\si{nm}$ and $v_{\rm col} \ll 50~\si{m.s^{-1}}$, and collisional deformation of dust aggregates due to rolling motion would be suppressed for large $\xi_{\rm crit}$ and/or small $v_{\rm col}$ cases.

\subsection{Initial condition}

As the initial condition, we use the two dust aggregates prepared by ballistic particle--cluster aggregation.
These aggregates are identical to those used in \citet{2022ApJ...933..144A}.
The number of particles constituting the target aggregate, $N_{\rm tar}$, is equal to that for the projectile aggregate, $N_{\rm pro}$, and $N_{\rm tar} = N_{\rm pro} = 50000$.
The total number of particles in a simulation, $N_{\rm tot}$, is $N_{\rm tot} = N_{\rm tar} + N_{\rm pro} = 10^{5}$.

Offset collisions of two dust aggregates is considered.
We use the normalized impact parameter, $B_{\rm off}$ \citep[see][]{2022ApJ...933..144A}.
The square of the normalized impact parameter ranges from ${B_{\rm off}}^{2} = 0$ to $1$ with an interval of $1/12$.
The collision velocity of two dust aggregates is set to $v_{\rm col} = 10^{( 0.1 i )}~\si{m.s^{-1}}$, where $i = 13$, $14$, \ldots, $20$.
We also change $\xi_{\rm crit}$ that controls the strength of interparticle rolling friction.
We set $\xi_{\rm crit} = 0.2 \times 2^{j}~\si{nm}$, where $j = 0$, $1$, \ldots, $6$; the parameter range of $\xi_{\rm crit}$ satisfies $0.2~\si{nm} \lesssim \xi_{\rm crit} \lesssim a_{0}$.

\section{Examples of collisional outcomes}

\begin{figure*}
\centering
\includegraphics[width=0.8\textwidth]{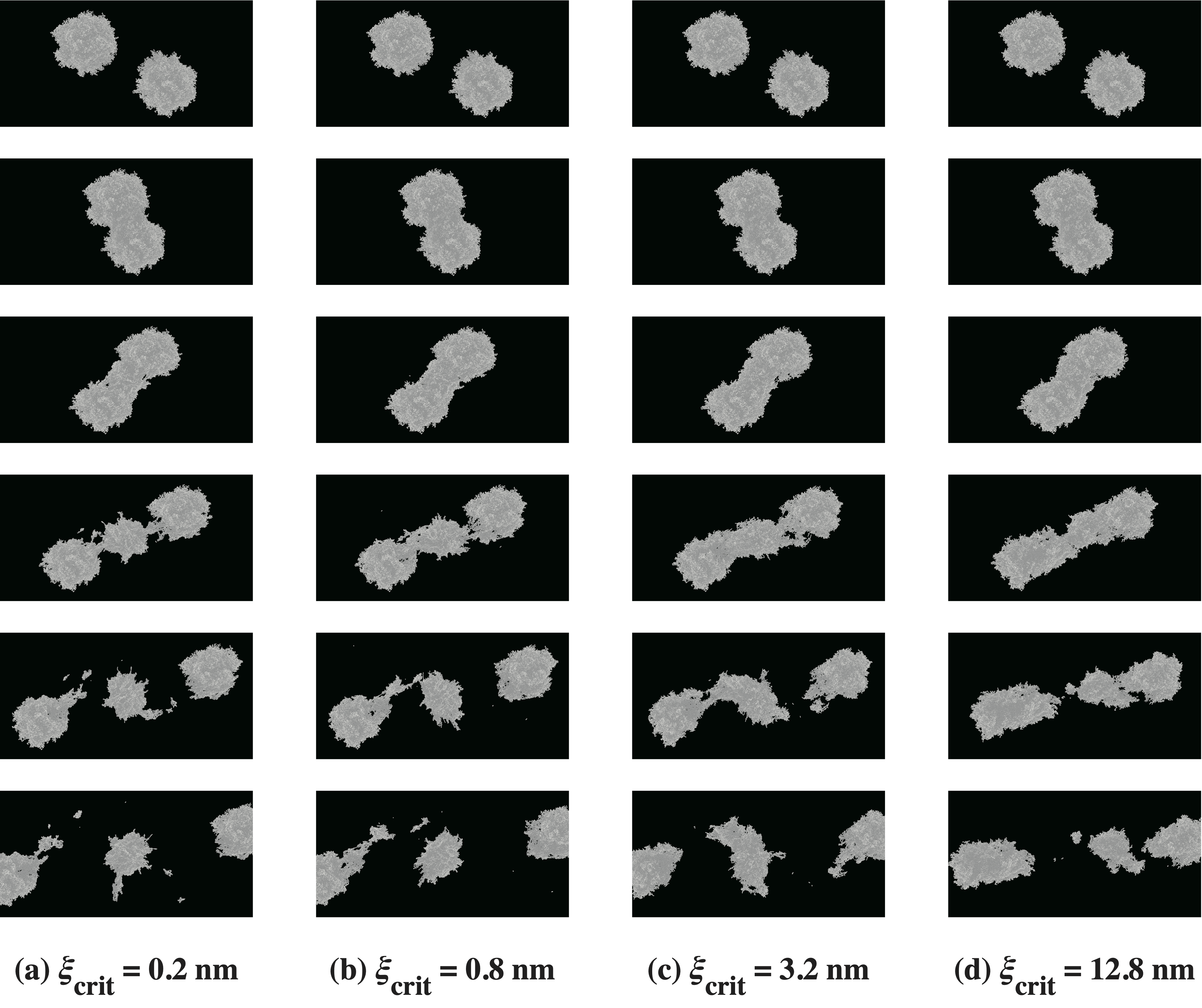}
\caption{
Snapshots of collisional outcomes.
Here we set $v_{\rm col} = 39.8~\si{m.s^{-1}}$ and ${B_{\rm off}}^{2} = 6/12$.
Panels (a)--(d) are time series of snapshots for $\xi_{\rm crit} = 0.2~\si{nm}$, $0.8~\si{nm}$, $3.2~\si{nm}$, and $12.8~\si{nm}$, respectively.
The time interval for each snapshot is $0.40~\si{\micro s}$.
}
\label{fig.snapshot}
\end{figure*}

Figure \ref{fig.snapshot} shows the snapshots of collisional outcomes.
Here we show the results for $v_{\rm col} = 39.8~\si{m.s^{-1}}$ and ${B_{\rm off}}^{2} = 6/12$. 
We found that the collisional behavior is similar among different settings of $\xi_{\rm crit}$ as shown in Figure \ref{fig.snapshot}.
We also analyzed the number of constituent particles in each fragment at the end of simulation, $t = 7.99~\si{\micro s}$.
The numbers of constituent particles in the three largest remnants are ${( N_{\rm lar}, N_{\rm 2nd}, N_{\rm 3rd} )} = {( 34042, 33299, 29827 )}$ for $\xi_{\rm crit} = 0.2~\si{nm}$, ${( 34170, 32457, 30940 )}$ for $\xi_{\rm crit} = 0.8~\si{nm}$, ${( 32571, 31598, 29338 )}$ for $\xi_{\rm crit} = 3.2~\si{nm}$, and ${( 48491, 31900, 19003 )}$ for $\xi_{\rm crit} = 12.8~\si{nm}$, respectively.
For these cases, most of particles are included in the three largest remnants.

\section{Growth efficiency}

\begin{figure*}
\centering
\includegraphics[width=0.48\textwidth]{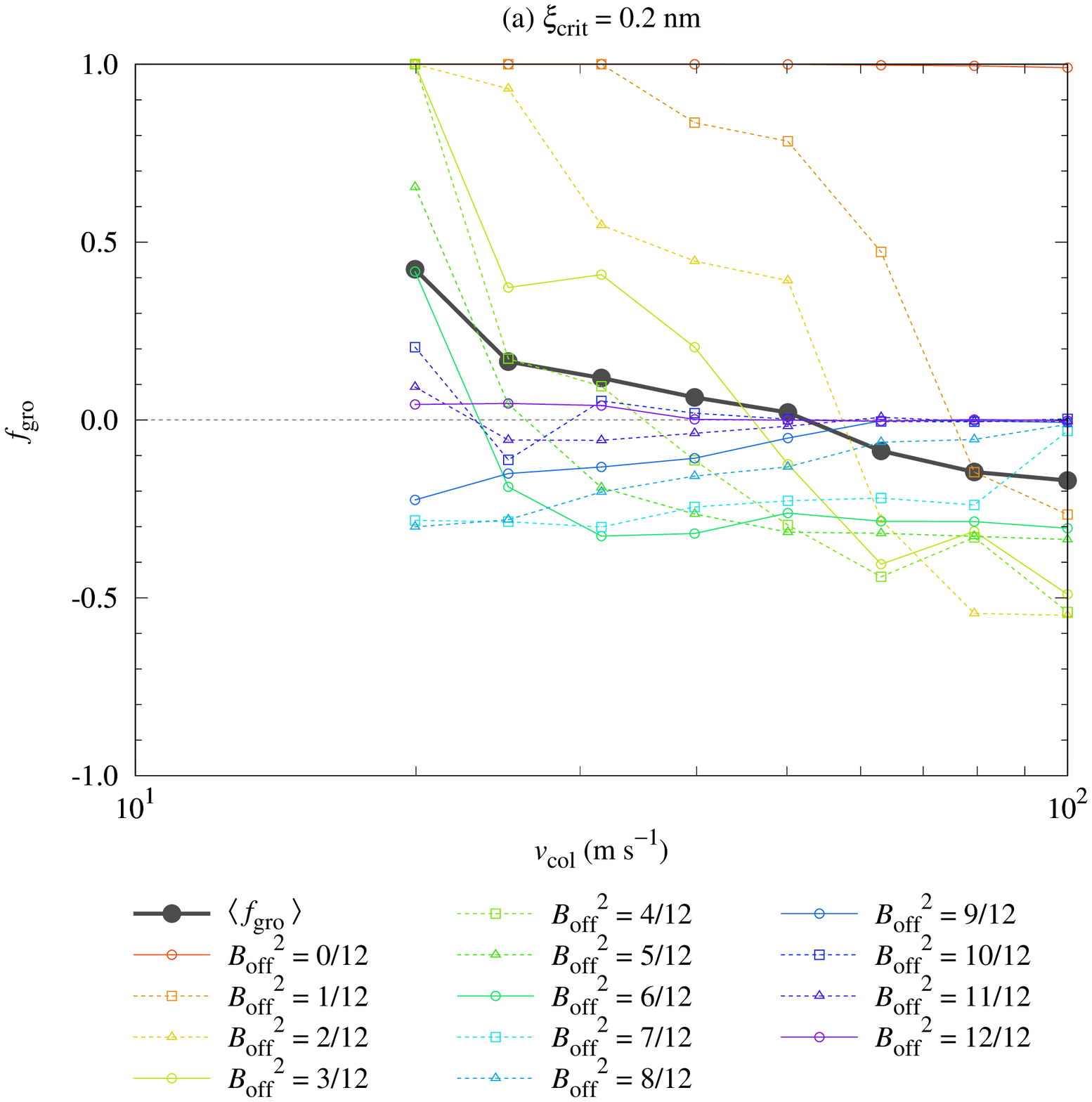}
\includegraphics[width=0.48\textwidth]{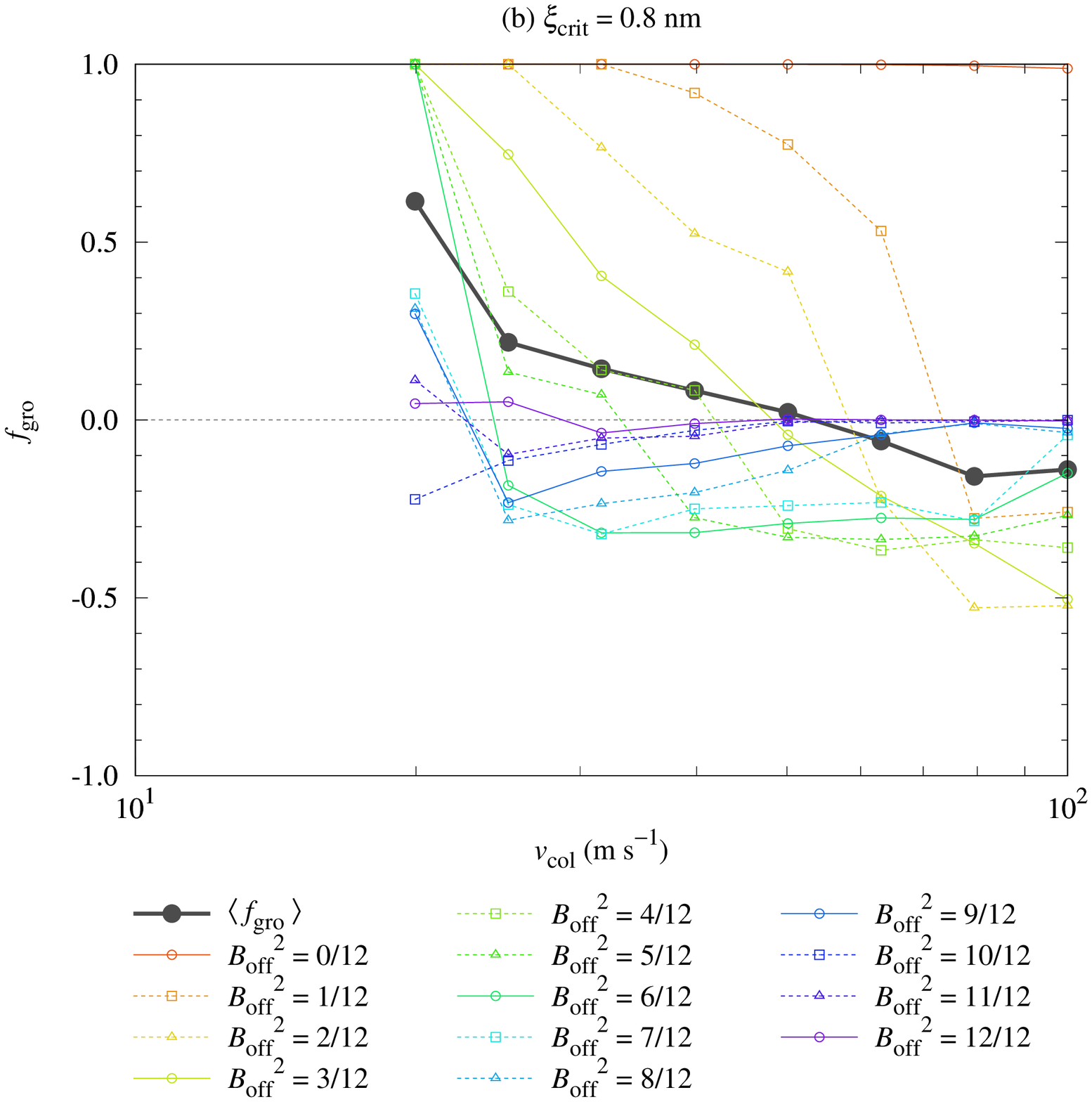}
\includegraphics[width=0.48\textwidth]{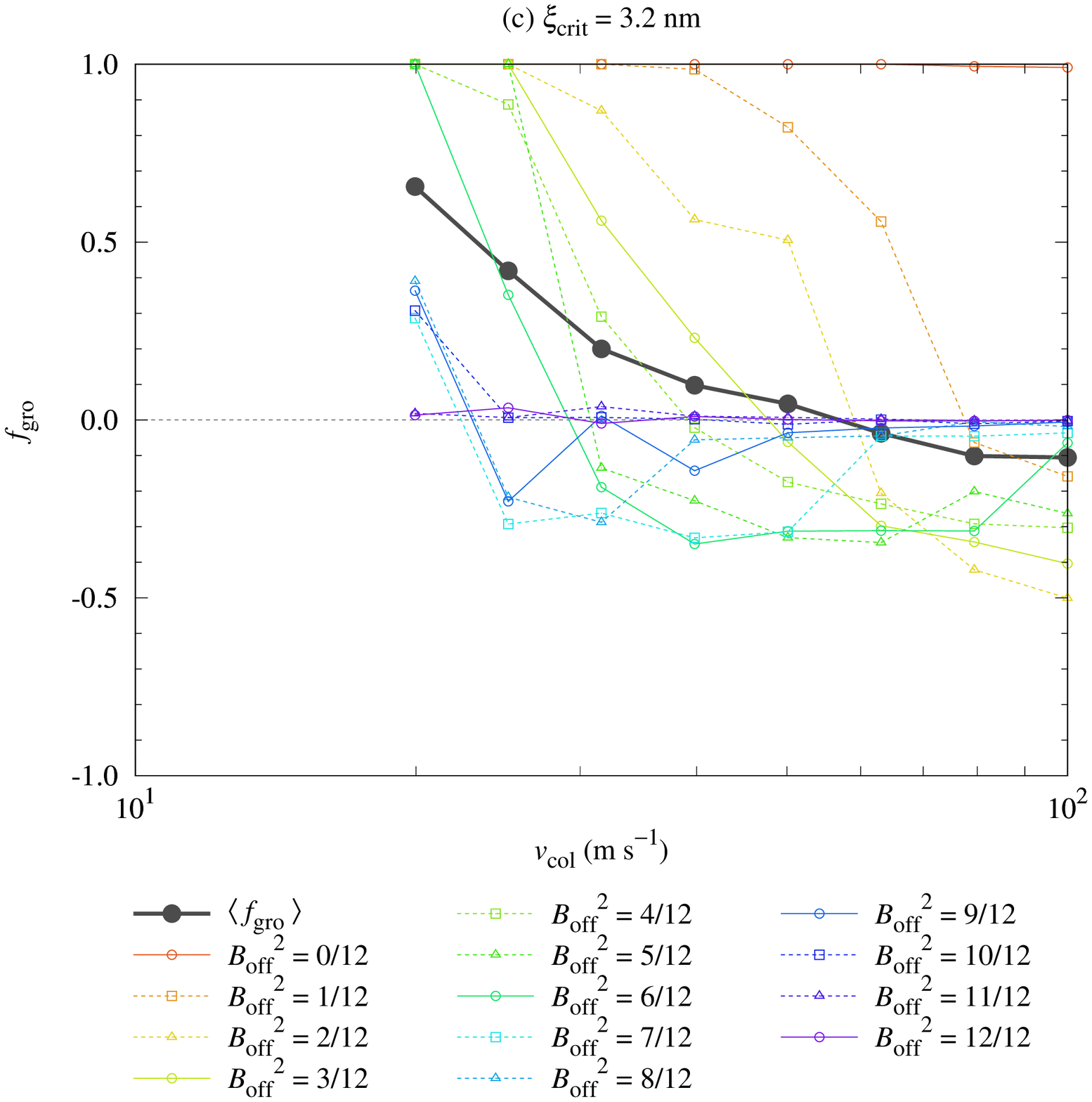}
\includegraphics[width=0.48\textwidth]{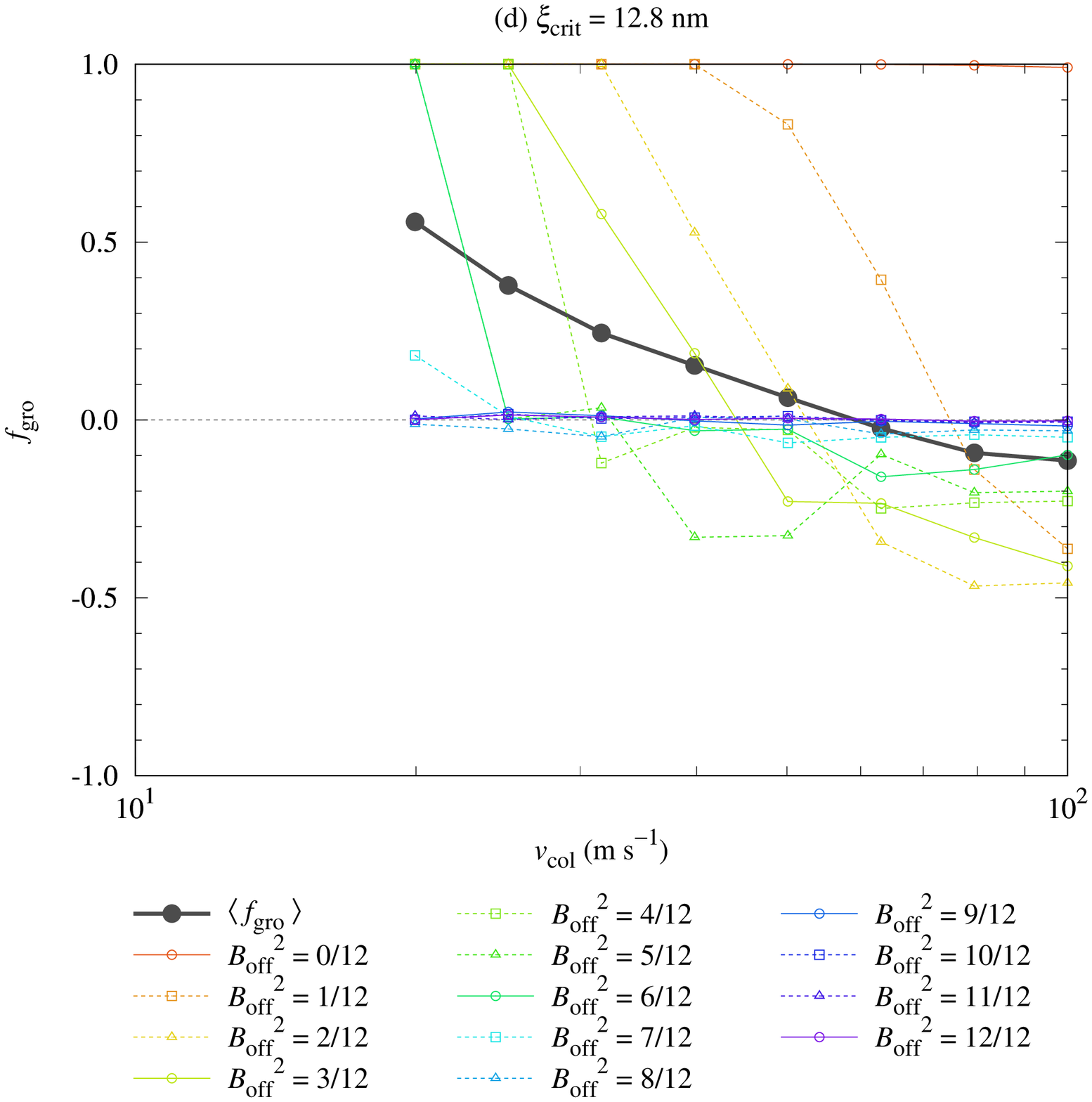}
\caption{
Collisional growth efficiency, $f_{\rm gro}$, for different settings of $\xi_{\rm crit}$, $B_{\rm off}$, and $v_{\rm col}$.
}
\label{fig.fgro}
\end{figure*}

We discuss the parameter dependence of the largest fragment formed after a collision.
The collisional growth efficiency, $f_{\rm gro}$, is defined as follows \citep[e.g.,][]{2013A&A...559A..62W}:
\begin{equation}
f_{\rm gro} = \frac{N_{\rm lar} - N_{\rm tar}}{N_{\rm pro}}.
\end{equation}
Figure \ref{fig.fgro} shows the collisional growth efficiency for different settings of $\xi_{\rm crit}$, $B_{\rm off}$, and $v_{\rm col}$.
Panels (a)--(d) show the results for $\xi_{\rm crit} = 0.2~\si{nm}$, $0.8~\si{nm}$, $3.2~\si{nm}$, and $12.8~\si{nm}$, respectively.
For head-on collisions with ${B_{\rm off}}^{2} = 0/12$, $f_{\rm gro} \simeq 1$ when $20~\si{m.s^{-1}} \le v_{\rm col} \le 100~\si{m.s^{-1}}$, and for grazing collisions with ${B_{\rm off}}^{2} = 12/12$, $f_{\rm gro} \simeq 0$.
These trends are consistent with those observed in previous studies \citep[e.g.,][]{2021ApJ...915...22H, 2022ApJ...933..144A}.

We note that \citet{2022MNRAS.509.5641S} proposed a model for collisions between porous dust aggregates based on their experimental results.
In their model, head-on collisions of porous aggregates result in bouncing when the collision velocity exceeds the sticking threshold, and the sticking threshold decreases with increasing $\xi_{\rm crit}$.
However, we have never seen bouncing behavior for head-on collisions even though $\xi_{\rm crit}$ is varied.
Although the real reason is unclear, this discrepancy might originate from the difference in radius and/or structure of aggregates; the radius of dust aggregates used in this study is approximately $7~\si{\micro m}$, while \citet{2022MNRAS.509.5641S} used cm-sized aggregates.

The impact parameter must vary with each collision event in protoplanetary disks. 
Then the average value of $f_{\rm gro}$ weighted over $B_{\rm off}$ would be useful to investigate whether dust aggregates can grow into larger aggregates in protoplanetary disks.
The average value of $f_{\rm gro}$ weighted over $B_{\rm off}$ is given by
\begin{equation}
{\langle f_{\rm gro} \rangle} \equiv \frac{1}{\pi} \int_{0}^{1} {\rm d}{B_{\rm off}}~2 \pi B_{\rm off} f_{\rm gro}.
\end{equation}
The gray lines in Figure \ref{fig.fgro} are $B_{\rm off}$-weighted collisional growth efficiency, ${\langle f_{\rm gro} \rangle}$.
As ${\langle f_{\rm gro} \rangle}$ decreases with increasing $v_{\rm col}$, the threshold collision velocity for the fragmentation of dust aggregates, $v_{\rm fra}$, is defined as the velocity where ${\langle f_{\rm gro} \rangle} = 0$.
We found that $v_{\rm fra}$ is nearly independent of $\xi_{\rm crit}$; $50~\si{m.s^{-1}} \le v_{\rm fra} \le 60~\si{m.s^{-1}}$ for $0.2~\si{nm} \le \xi_{\rm crit} \le 12.8~\si{nm}$.

\begin{figure*}
\centering
\includegraphics[width=0.48\textwidth]{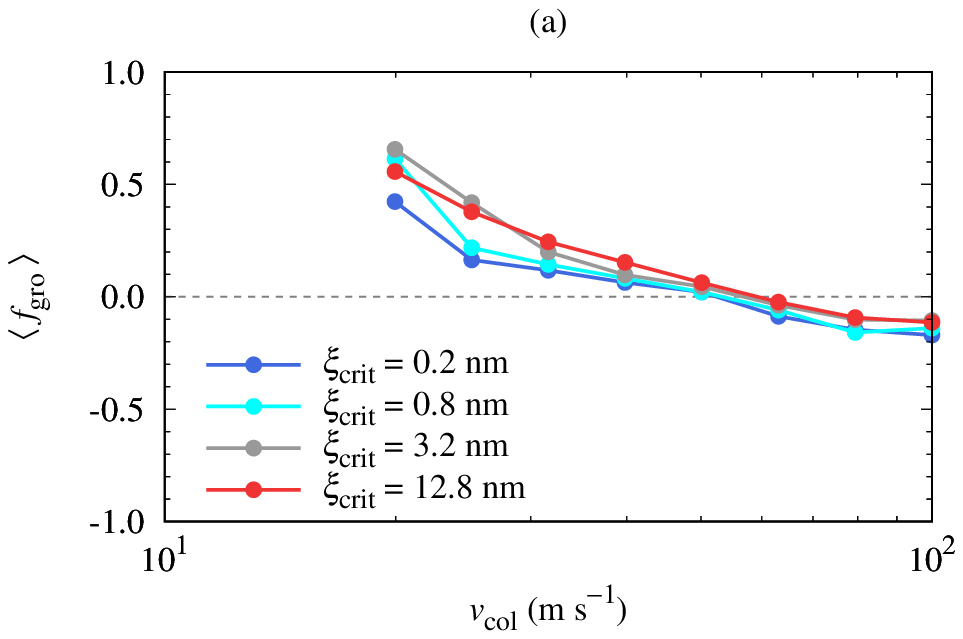}
\includegraphics[width=0.48\textwidth]{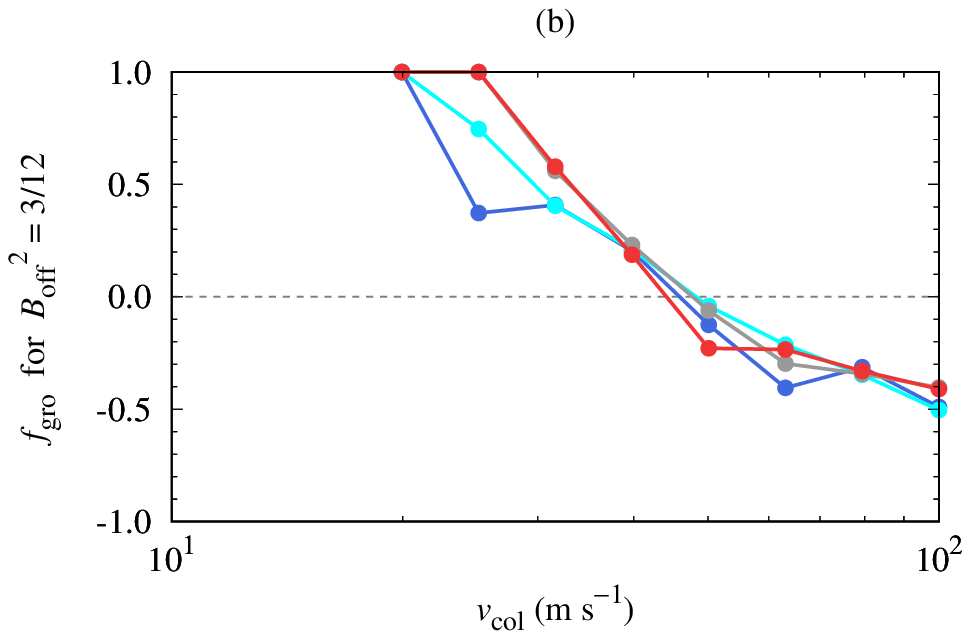}
\includegraphics[width=0.48\textwidth]{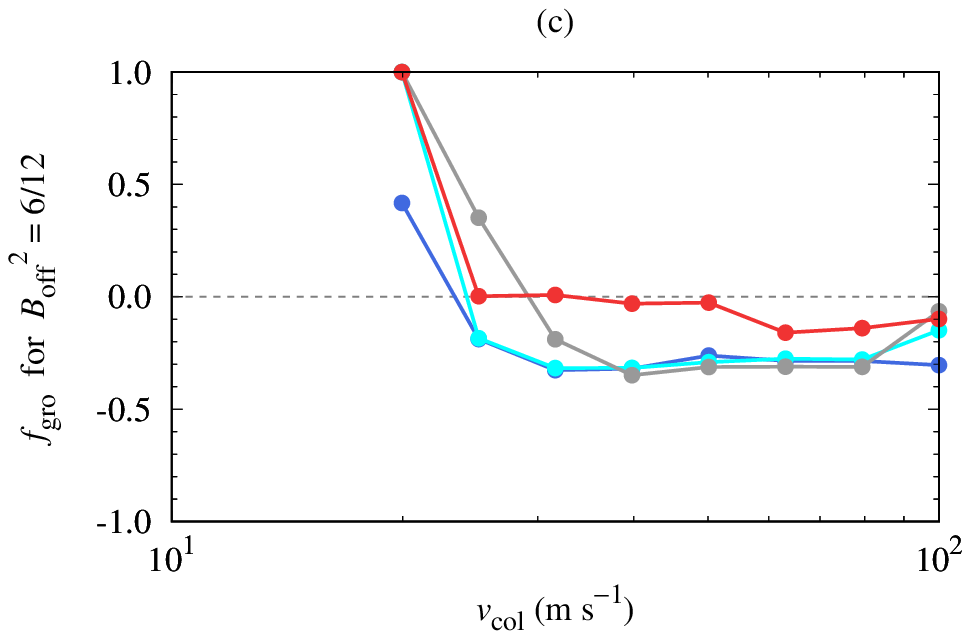}
\includegraphics[width=0.48\textwidth]{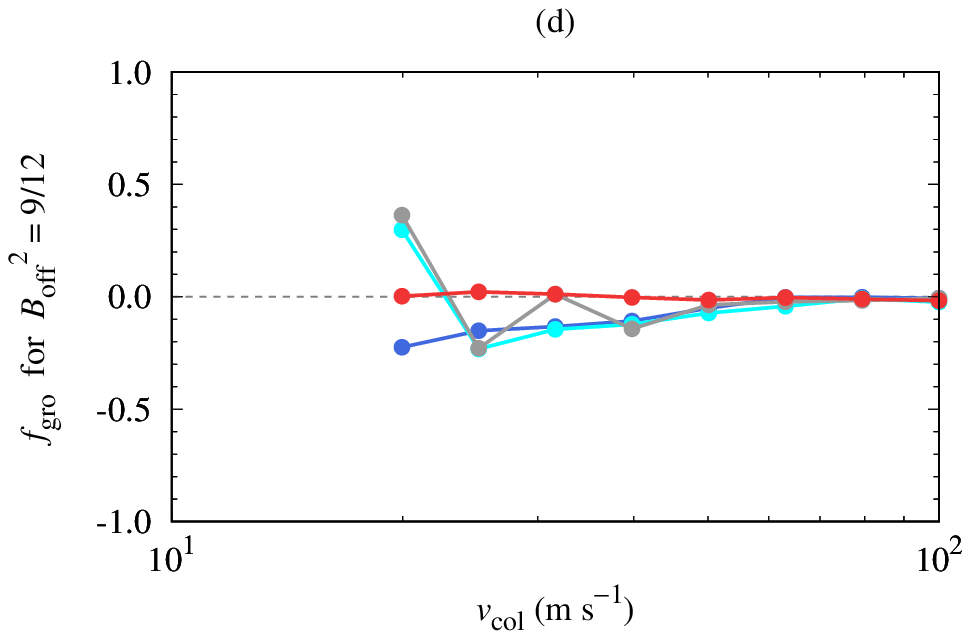}
\caption{
$B_{\rm off}$-weighted collisional growth efficiency (Panel (a)) and collisional growth efficiency for oblique collisions (${B_{\rm off}}^{2} = 3/12$ for Panel (b), ${B_{\rm off}}^{2} = 6/12$ for Panel (c), and ${B_{\rm off}}^{2} = 9/12$ for Panel (d)). 
}
\label{fig.fgro_ave}
\end{figure*}

Figure \ref{fig.fgro_ave}(a) shows the $B_{\rm off}$-weighted collisional growth efficiency.
Although ${\langle f_{\rm gro} \rangle}$ slightly depends on $\xi_{\rm crit}$ when we set $20~\si{m.s^{-1}} \le v_{\rm col} \le 30~\si{m.s^{-1}}$, $v_{\rm fra}$ is nearly independent of $\xi_{\rm crit}$ in our simulations.
Figures \ref{fig.fgro_ave}(b)--\ref{fig.fgro_ave}(d) show the collisional growth efficiency for oblique collisions.
As shown in Figures \ref{fig.fgro_ave}(c) and \ref{fig.fgro_ave}(d), for grazing collisions with ${B_{\rm off}}^{2} \ge 6/12$, $f_{\rm gro}$ is nearly independent of $v_{\rm col}$ around $v_{\rm fra}$.
In contrast, Figures \ref{fig.fgro_ave}(b) exhibits a clear dependence of $f_{\rm gro}$ on $v_{\rm col}$ around $v_{\rm fra}$ if we focus on oblique collisions with ${B_{\rm off}}^{2} \simeq 3/12$.
Therefore, we can imagine that the energy dissipation for oblique collisions with ${B_{\rm off}}^{2} \simeq 3/12$ and its dependence on $v_{\rm col}$ would be the key to unveiling the $\xi_{\rm crit}$-dependence of ${\langle f_{\rm gro} \rangle}$ and $v_{\rm fra}$.

\section{Energy dissipation}

\begin{figure*}
\centering
\includegraphics[width=0.48\textwidth]{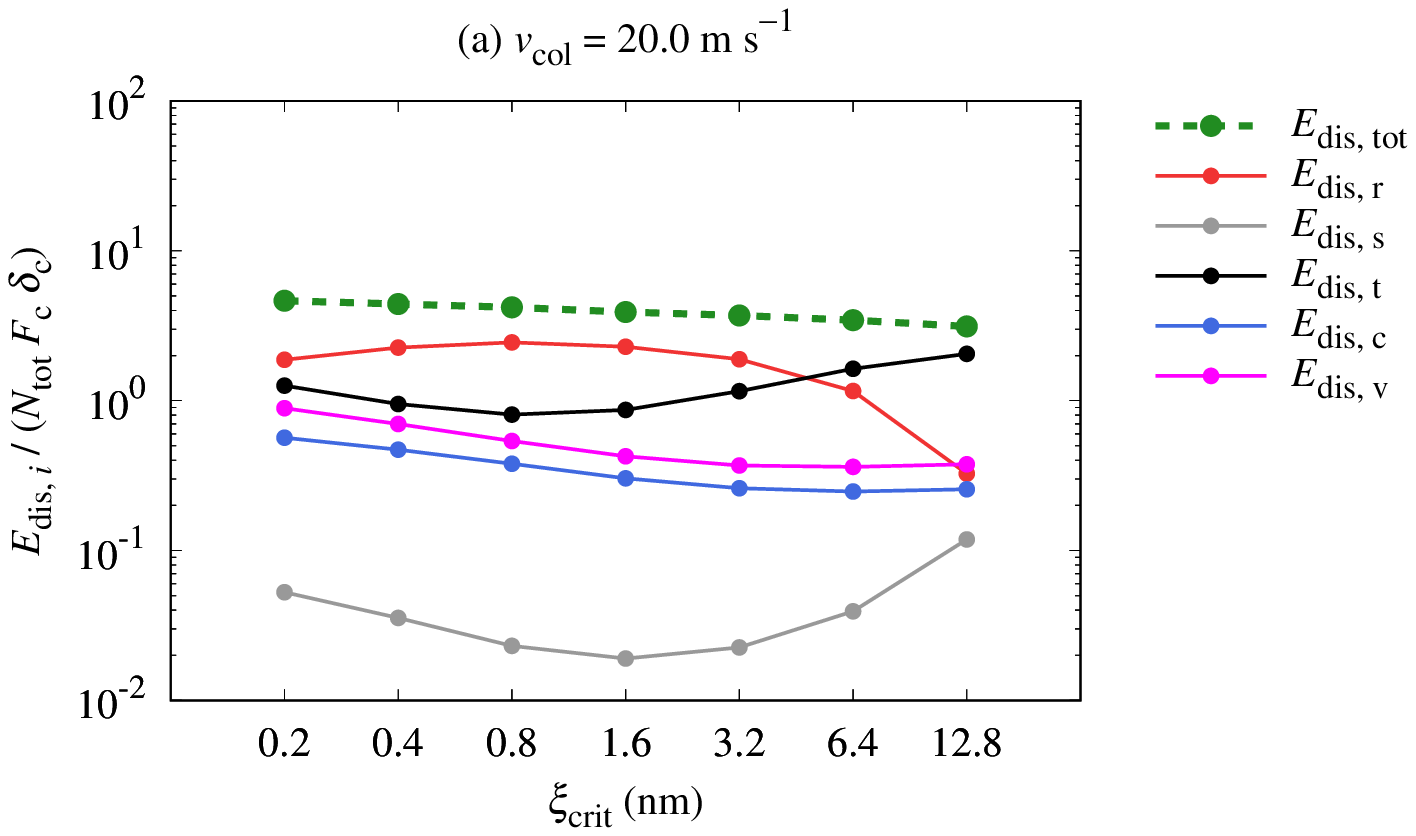}
\includegraphics[width=0.48\textwidth]{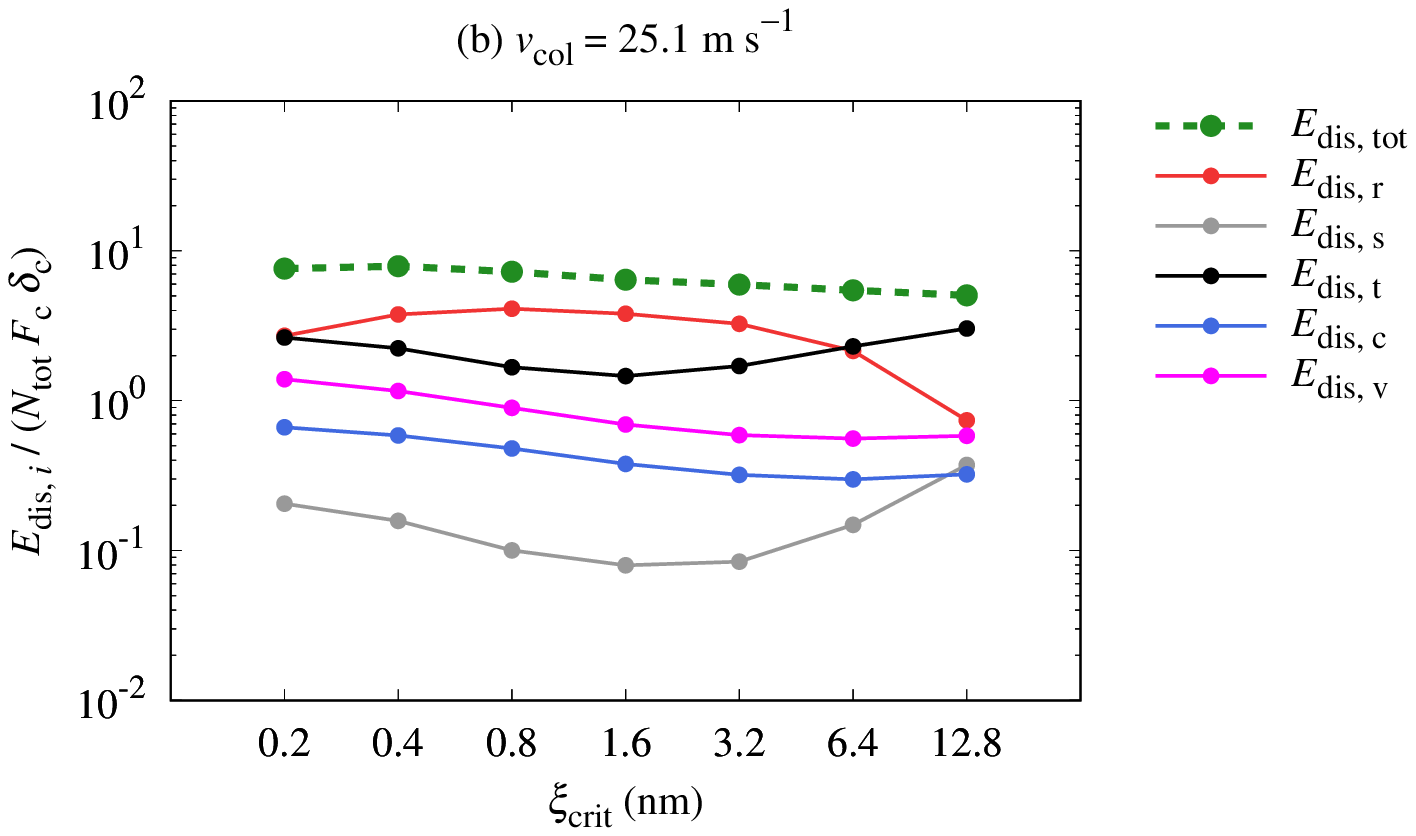}
\includegraphics[width=0.48\textwidth]{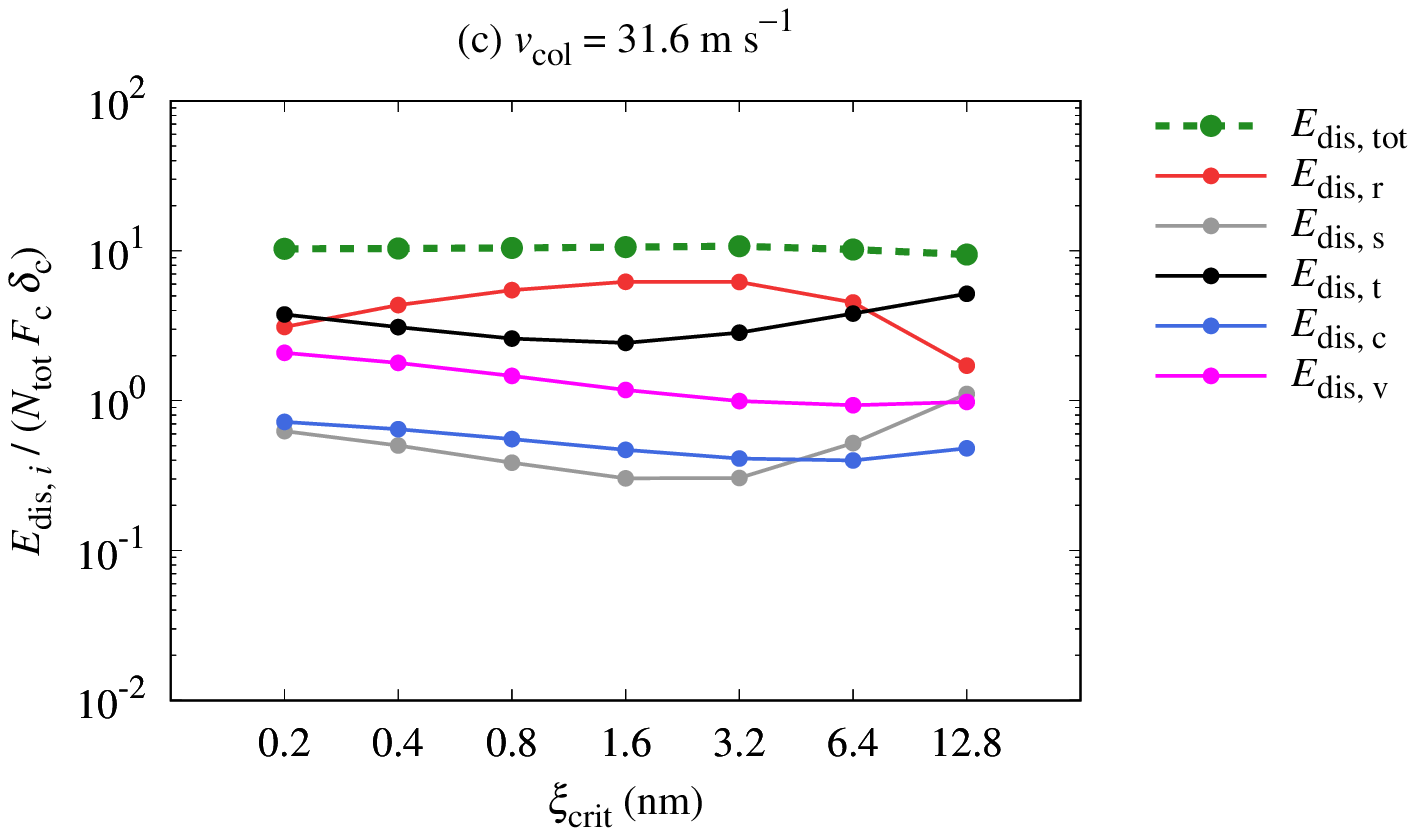}
\includegraphics[width=0.48\textwidth]{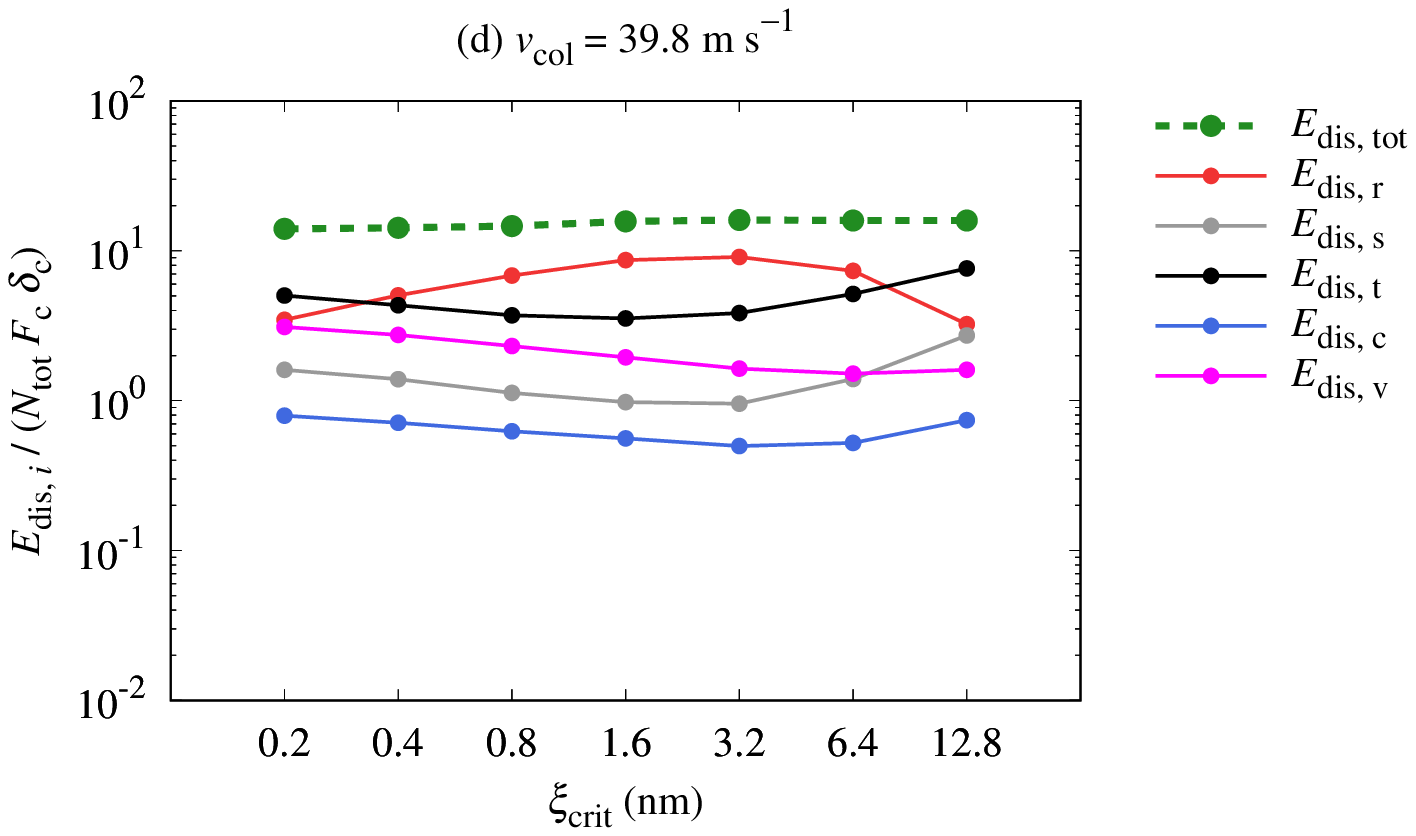}
\includegraphics[width=0.48\textwidth]{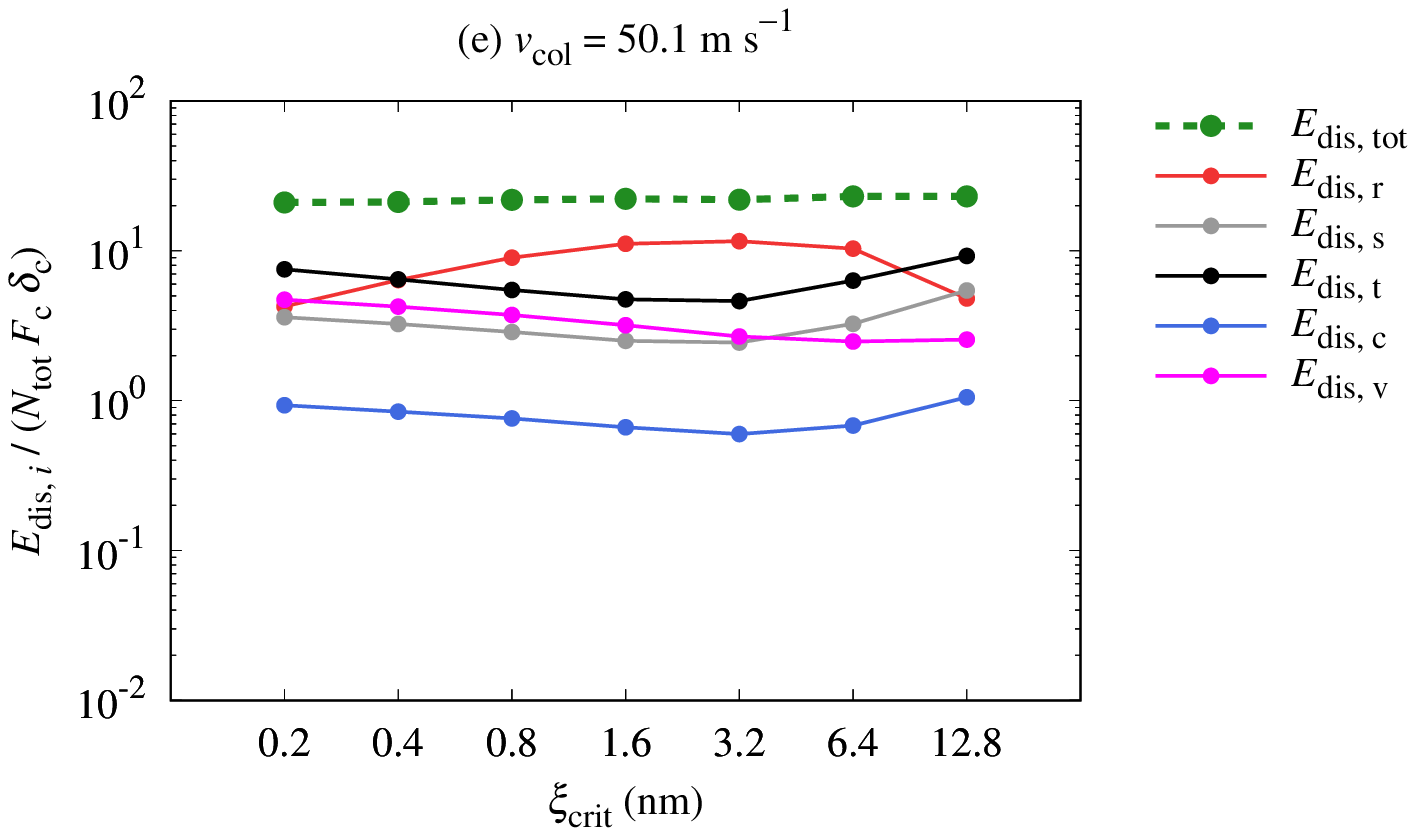}
\includegraphics[width=0.48\textwidth]{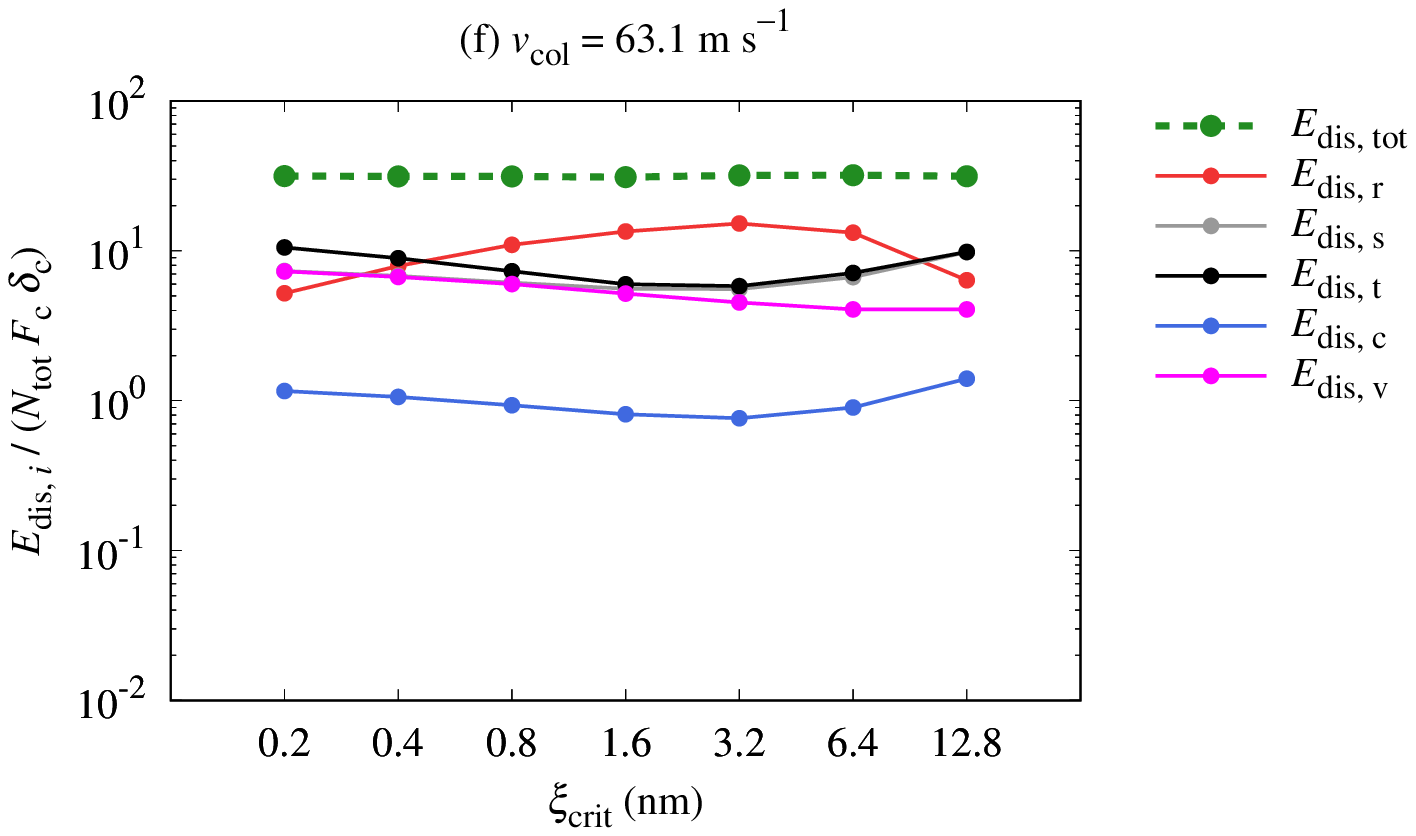}
\includegraphics[width=0.48\textwidth]{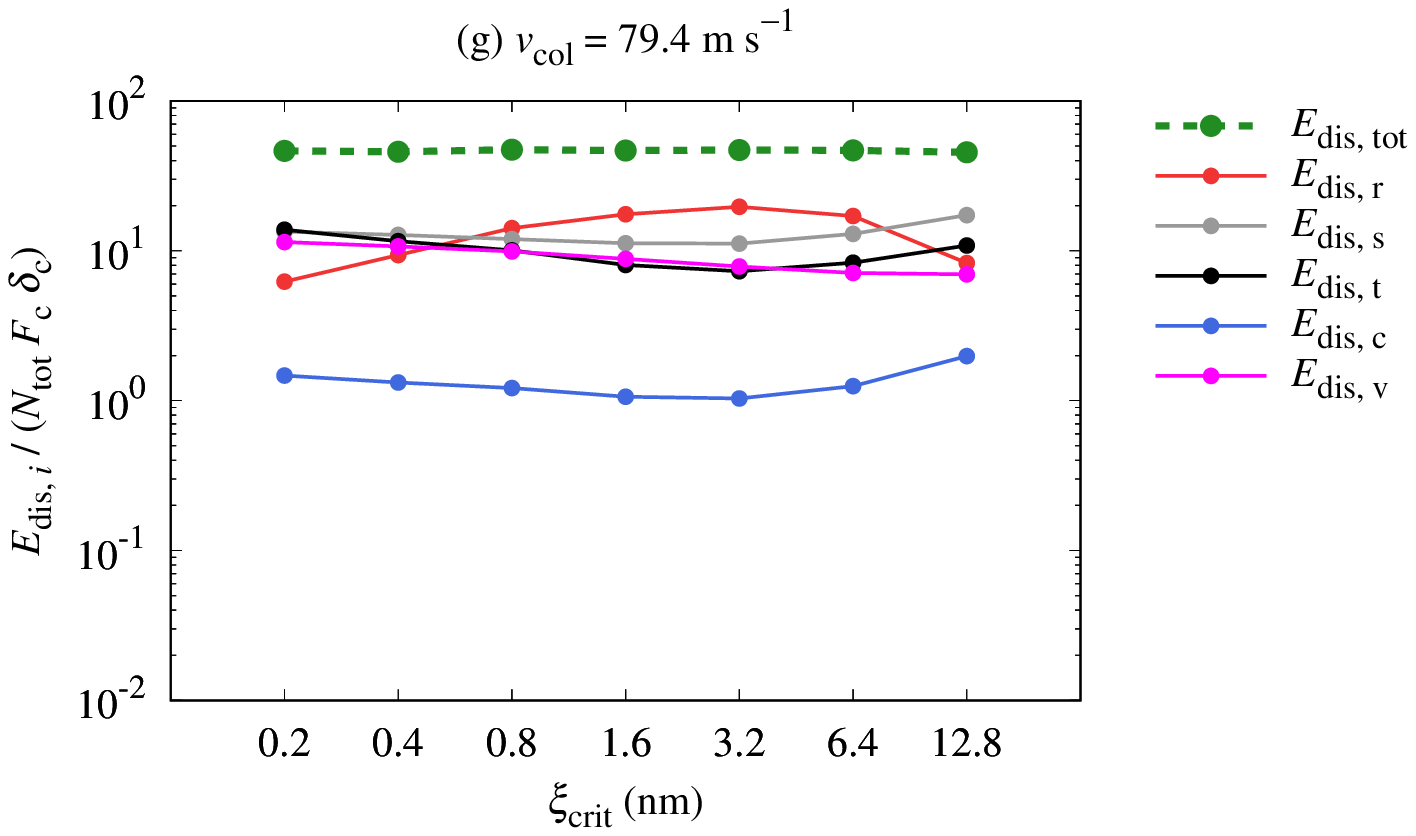}
\includegraphics[width=0.48\textwidth]{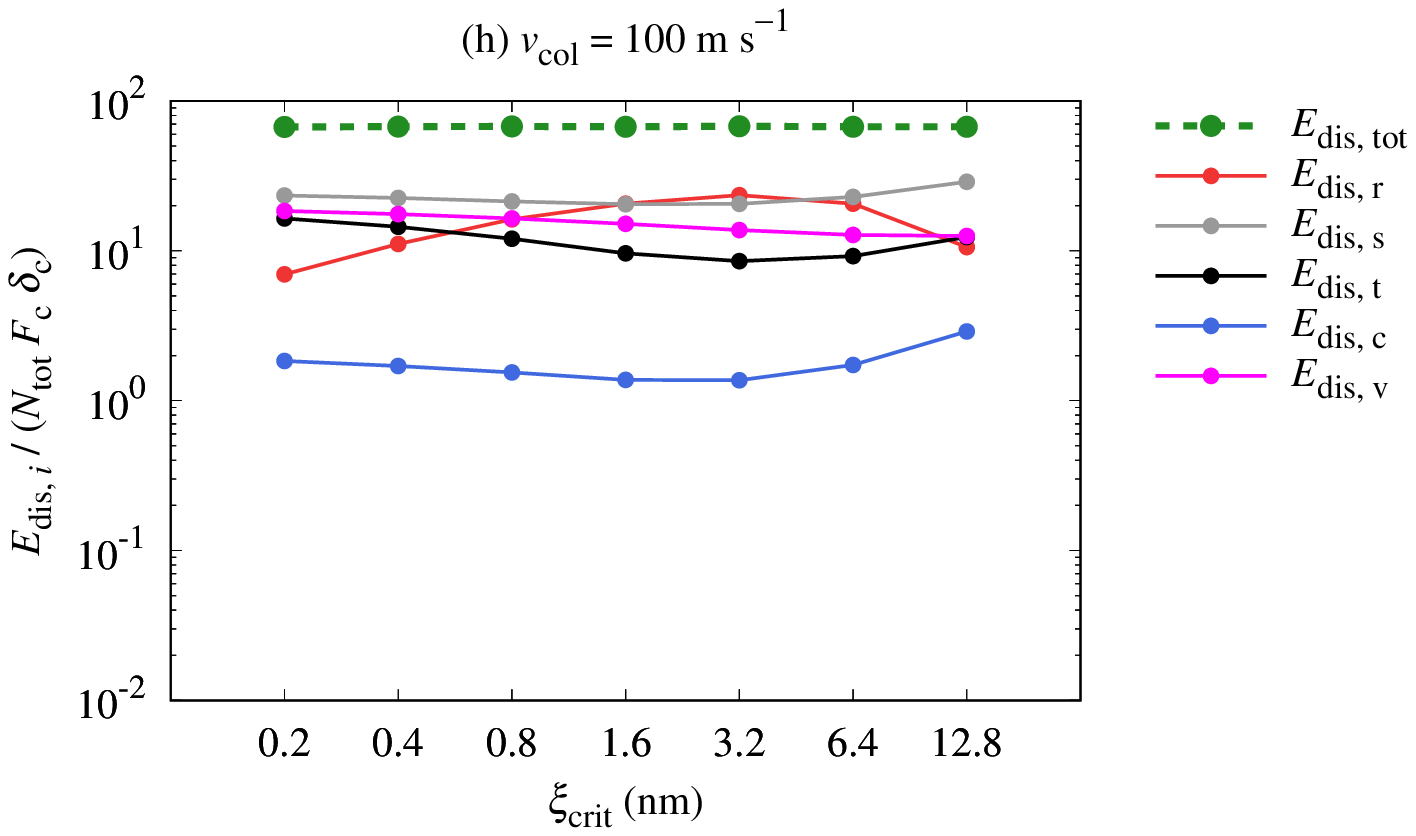}
\caption{
Total energy dissipation due to particle interactions from the start to the end of the simulations.
The total energy dissipation due to all particle interactions, $E_{\rm dis, tot}$, is given by $E_{\rm dis, tot} \equiv E_{\rm dis, r} + E_{\rm dis, s} + E_{\rm dis, t} + E_{\rm dis, c} + E_{\rm dis, v}$, where $E_{\rm dis, r}$, $E_{\rm dis, s}$, $E_{\rm dis, t}$, $E_{\rm dis, c}$, and $E_{\rm dis, v}$ are the energy dissipations due to rolling friction, sliding friction, twisting friction, connection and disconnection of particles, and viscous drag force, respectively.
Here we set ${B_{\rm off}}^{2} = 3/12$.
}
\label{fig.edis}
\end{figure*}

We check the energy dissipation due to particle interactions from the start to the end of the simulations.
Figure \ref{fig.edis} shows the energy dissipation due to rolling friction ($E_{\rm dis, r}$), sliding friction ($E_{\rm dis, s}$), twisting friction ($E_{\rm dis, t}$), connection and disconnection of particles ($E_{\rm dis, c}$), and viscous drag force ($E_{\rm dis, v}$).
We also show the total energy dissipation due to all particle interactions, $E_{\rm dis, tot} \equiv E_{\rm dis, r} + E_{\rm dis, s} + E_{\rm dis, t} + E_{\rm dis, c} + E_{\rm dis, v}$.
Here we set ${B_{\rm off}}^{2} = 3/12$ and investigated the dependence on $\xi_{\rm crit}$ and $v_{\rm col}$.
Results for ${B_{\rm off}}^{2} = 6/12$ and $9/12$ are shown in Appendix \ref{app.edis} as a reference (see Figures \ref{fig.edis6} and \ref{fig.edis9}).
The particle interaction energies for interparticle motions (e.g., $E_{\rm r, crit}$ and $E_{\rm roll}$) are summarized in Table 2 of \citet{2022ApJ...933..144A}.

We found that $E_{\rm dis, r}$ does not monotonically increases with $\xi_{\rm crit}$; there is a maximum value of $E_{\rm dis, r}$.
It is clear that $E_{\rm dis, r} \to 0$ when $\xi_{\rm crit} \to 0$ because $E_{\rm roll}$ becomes zero.
It is also natural that $E_{\rm dis, r} \to 0$ when $\xi_{\rm crit} \to \infty$; this is because the critical energy required to start rolling becomes $E_{\rm r, crit} \to \infty$.
Although it depends on $v_{\rm col}$, $E_{\rm dis, r}$ takes the maximum around $\xi_{\rm crit} \simeq 3.2~\si{nm}$ for $v_{\rm col} \sim v_{\rm fra}$.
As the kinetic energy per one particle, $E_{\rm kin, 1}$, cannot overcome $E_{\rm roll}$ for $v_{\rm col} \sim v_{\rm fra}$ and $\xi_{\rm crit} \gg 3.2~\si{nm}$, the decrease of $E_{\rm dis, r}$ for large $\xi_{\rm crit}$ would be consistent with the order-of-magnitude estimate in Section \ref{sec.roll}.

In contrast to $E_{\rm dis, r}$, $E_{\rm dis, s}$ and $E_{\rm dis, t}$ take large values for both lower and upper limits of $\xi_{\rm crit}$ (i.e., $0.2~\si{nm}$ and $12.8~\si{nm}$, respectively). 
For low-speed collisions with $v_{\rm col} \lesssim v_{\rm fra}$, $E_{\rm dis, t}$ plays the major role in energy dissipation when $\xi_{\rm crit}$ is set to be close to lower or upper limits.
For high-speed collisions with $v_{\rm col} \gtrsim v_{\rm fra}$, $E_{\rm dis, s}$ plays the major role in energy dissipation when $\xi_{\rm crit}$ is set to be close to lower or upper limits.
Conversely, $E_{\rm dis, r}$ is the strongest energy dissipation mechanism when we choose a moderate $\xi_{\rm crit}$ in the range of $0.8$--$3.2~\si{nm}$ for submicron-sized ice particles.

The strong dependence of $E_{\rm dis, s}$ on $v_{\rm col}$ is discussed in Section 5.2 of \citet{2022ApJ...933..144A}.
They claimed that the strong dependence could be originated from the magnitude relation between $E_{\rm slide}$ and $E_{\rm kin, 1}$, where $E_{\rm slide}$ is the the energy needed to slide a particle by $\pi / 2$ radian around its contact point.

Interestingly, the total energy dissipation due to all particle interactions, $E_{\rm dis, tot}$, is nearly independent of $\xi_{\rm crit}$.\footnote{
Although the dependence of $E_{\rm dis, tot}$ on $\xi_{\rm crit}$ is weak even for small $v_{\rm col}$, we acknowledge that $E_{\rm dis, tot}$ decreases with increasing $\xi_{\rm crit}$ when $v_{\rm col} \ll v_{\rm fra}$, as shown in Figures \ref{fig.edis}(a) and \ref{fig.edis}(b).
}
In other words, our numerical results suggest that three mechanisms of tangential frictions could complement one another.
The weak dependence of $E_{\rm dis, tot}$ on $\xi_{\rm crit}$ might be the reason why ${\langle f_{\rm gro} \rangle}$ and $v_{\rm fra}$ are not strong functions of $\xi_{\rm crit}$.
As we prepared initial dust aggregates via sequential hit-and-stick collisions between an aggregate and multiple single particles \citep[e.g.,][]{1992A&A...262..315M}, they are highly porous and have only two interparticle contacts per particle in average.
Therefore, we can expect that there are higher degrees of freedom for deformation of dust aggregates, and all three tangential motions (rolling, sliding, and twisting) can potentially play a prominent role in energy dissipation if the strengths of springs and their critical displacements have moderate values.
We note that ${\langle f_{\rm gro} \rangle}$ and $v_{\rm fra}$ considerably decrease when we perform numerical simulations of collisions between aggregates made of cohesive but frictionless spheres (Arakawa et al., in preparation).
Although what controls the balance of energy dissipation via tangential frictions is still unclear, our results highlight the importance of three tangential motions on the collisional outcomes of porous dust aggregates made of submicron-sized particles.

\section{Summary}

The pairwise collisional growth of dust aggregates consisting submicron-sized grains is the first step of the planet formation, and understanding the collisional behavior of dust aggregates is therefore essential.
It is known that the main energy dissipation mechanisms are the tangential frictions between particles in contact, namely, rolling, sliding, and twisting \citep[e.g.,][]{2022ApJ...933..144A}.
However, there is a large uncertainty for the strength of rolling friction, and the dependence of the collisional growth condition on the strength of rolling friction was poorly understood.

In this study, we systematically performed numerical simulations of collisions between two equal-mass dust aggregates made of sibmicron-sized ice spheres.
We changed the strength of interparticle rolling friction, which is controlled by the critical rolling displacement, $\xi_{\rm crit}$.
As a results, we found that the threshold collision velocity for the fragmentation of dust aggregates is nearly independent of $\xi_{\rm crit}$ when we consider oblique collisions (Figures \ref{fig.fgro} and \ref{fig.fgro_ave}).

We checked the total energy dissipation due to particle interactions from the start to the end of the simulation (Figure \ref{fig.edis}).
We found that the total energy dissipation due to all particle interactions, $E_{\rm dis, tot}$, is nearly independent of $\xi_{\rm crit}$.
The energy dissipation due to rolling friction, $E_{\rm dis, r}$, takes the maximum when we chose a moderate $\xi_{\rm crit}$, while $E_{\rm dis, r}$ becomes small if $\xi_{\rm crit}$ is set to be close to lower or upper limits ($0.2~\si{nm}$ and $12.8~\si{nm}$, respectively).
Conversely, the energy dissipations due to sliding friction, $E_{\rm dis, s}$, and twisting friction, $E_{\rm dis, t}$, become large at both lower and upper limits of $\xi_{\rm crit}$.
Therefore, our numerical results indicate that three mechanisms of tangential frictions could complement one another.

Numerical simulations of granular matter are performed in many fields of science and engineering.
We note, however, that the particle interaction models differ among numerical codes.
Although the majority of these codes do not include the particle interaction for twisting motion, our simulations showed that twisting friction plays an important role when the collision velocity of dust aggregates, $v_{\rm col}$, is not significantly larger than the threshold collision velocity for the fragmentation, $v_{\rm fra}$.
Therefore, the detail of particle interaction models would be crucial for investigating the colliisonal behavior of dust aggregates from numerical simulations.

We acknowledge that plastic deformation, melting, and fragmentation of constituent particles are not considered in this study.
These effects might be important for high-speed collisions as shown in molecular dynamics simulations \citep[e.g.,][]{2020Icar..35213996N, 2022ApJ...925..173N}.
The angular velocity dependence of the strength of tangential frictions is also ignored in our simulations.
Although our results revealed that $v_{\rm fra}$ does not strongly depend on the strength of rolling friction alone, the collisional behavior might be affected when the strengths of all three tangential frictions increase/decrease simultaneously.
Future studies on the particle interactions from laboratory experiments and molecular dynamics simulations would be essential.


\section*{acknowledgments}
Numerical computations were carried out on PC cluster at CfCA, NAOJ.
H.T.\ and E.K.\ were supported by JSPS KAKENHI grant No.\ 18H05438.
This work was supported by the Publications Committee of NAOJ.

%

\appendix

\section{Total energy dissipation due to particle interactions}
\label{app.edis}

\begin{figure*}
\centering
\includegraphics[width=0.48\textwidth]{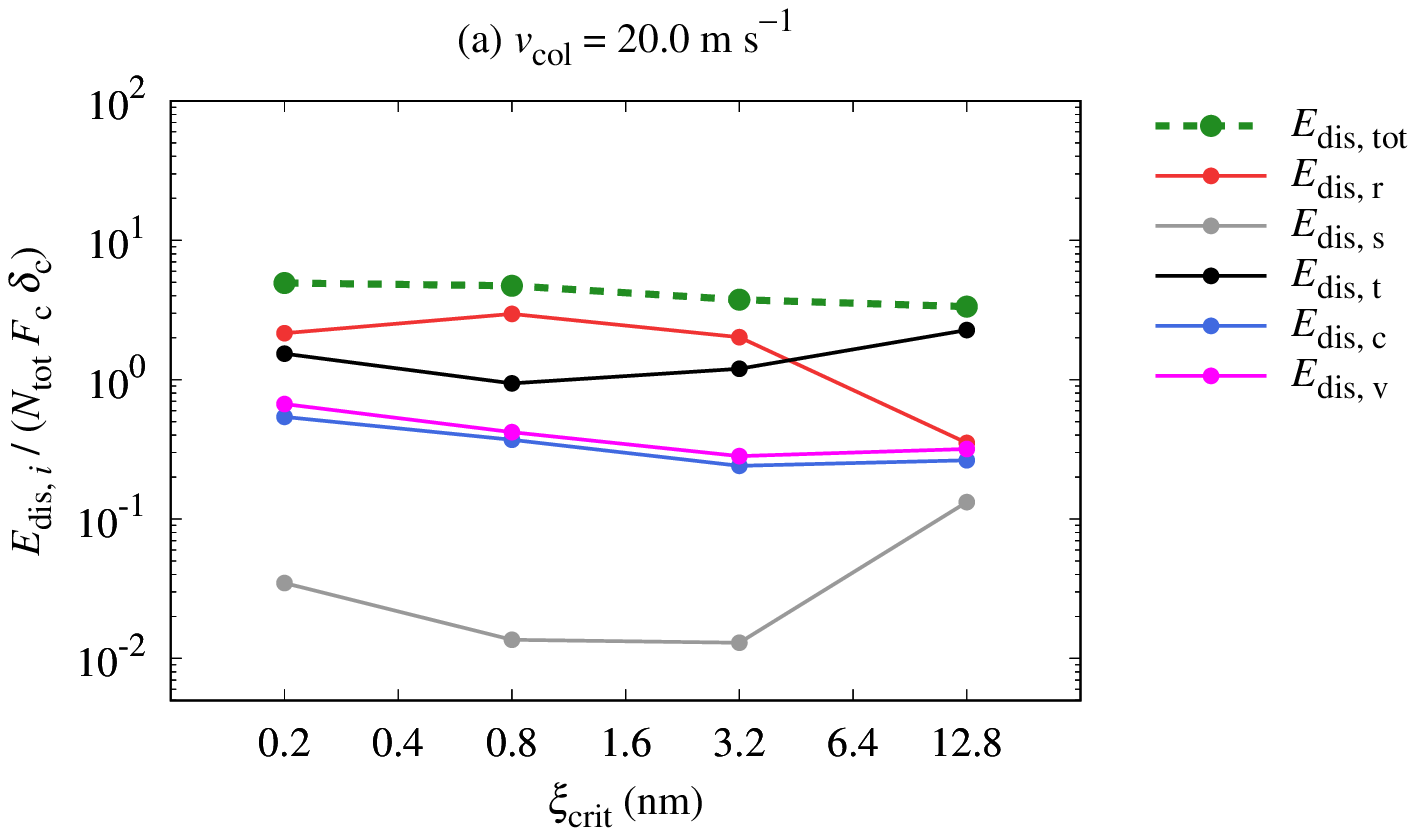}
\includegraphics[width=0.48\textwidth]{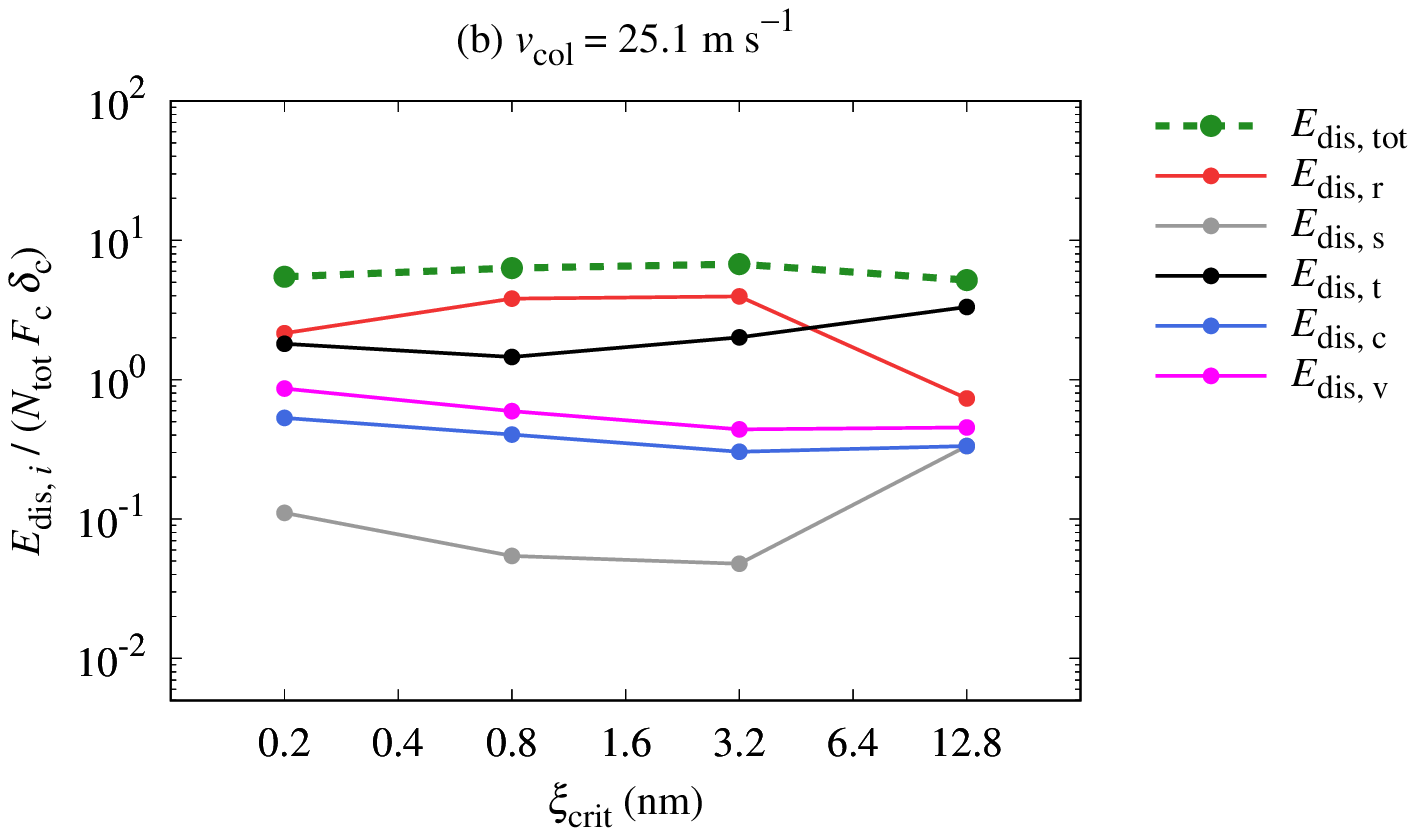}
\includegraphics[width=0.48\textwidth]{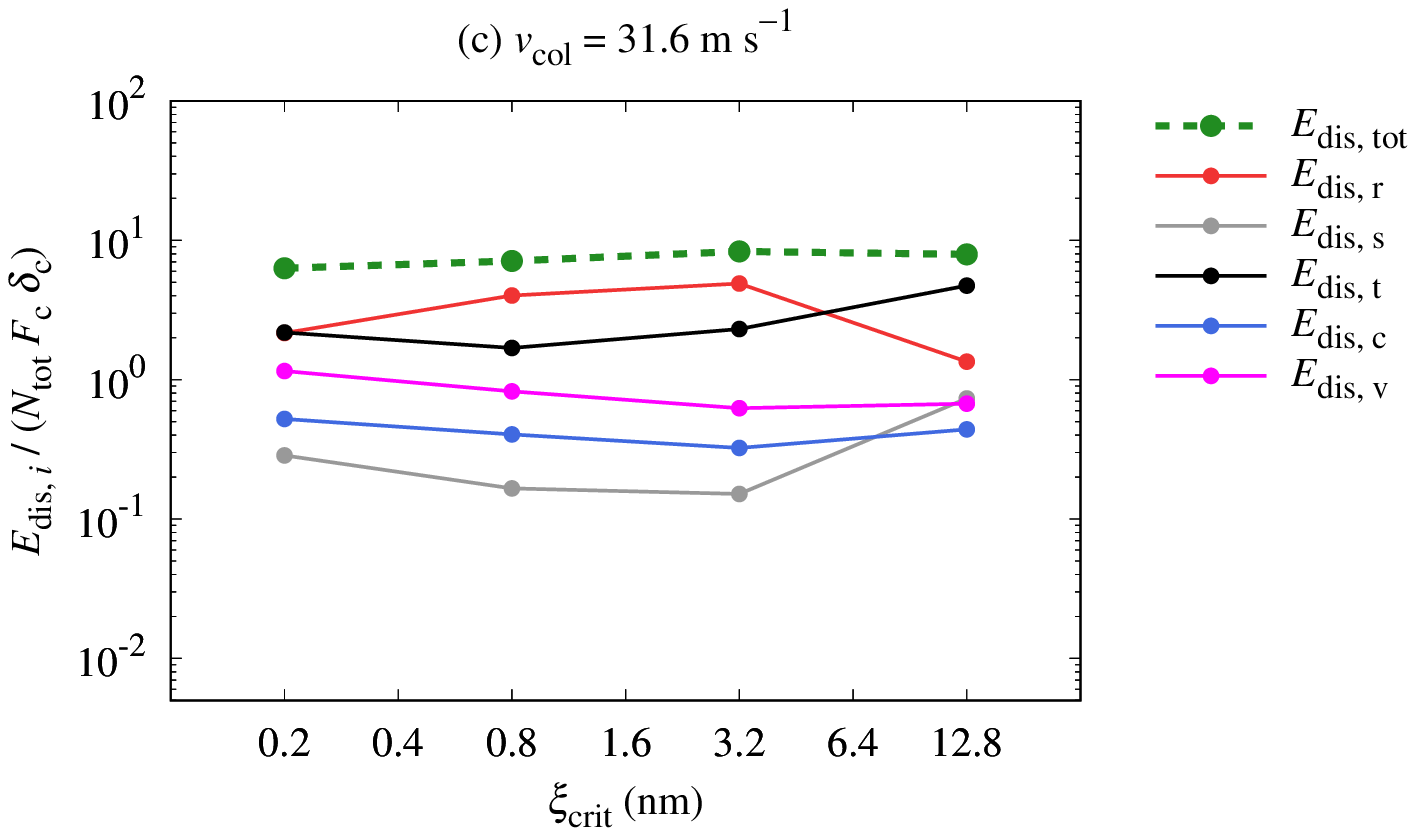}
\includegraphics[width=0.48\textwidth]{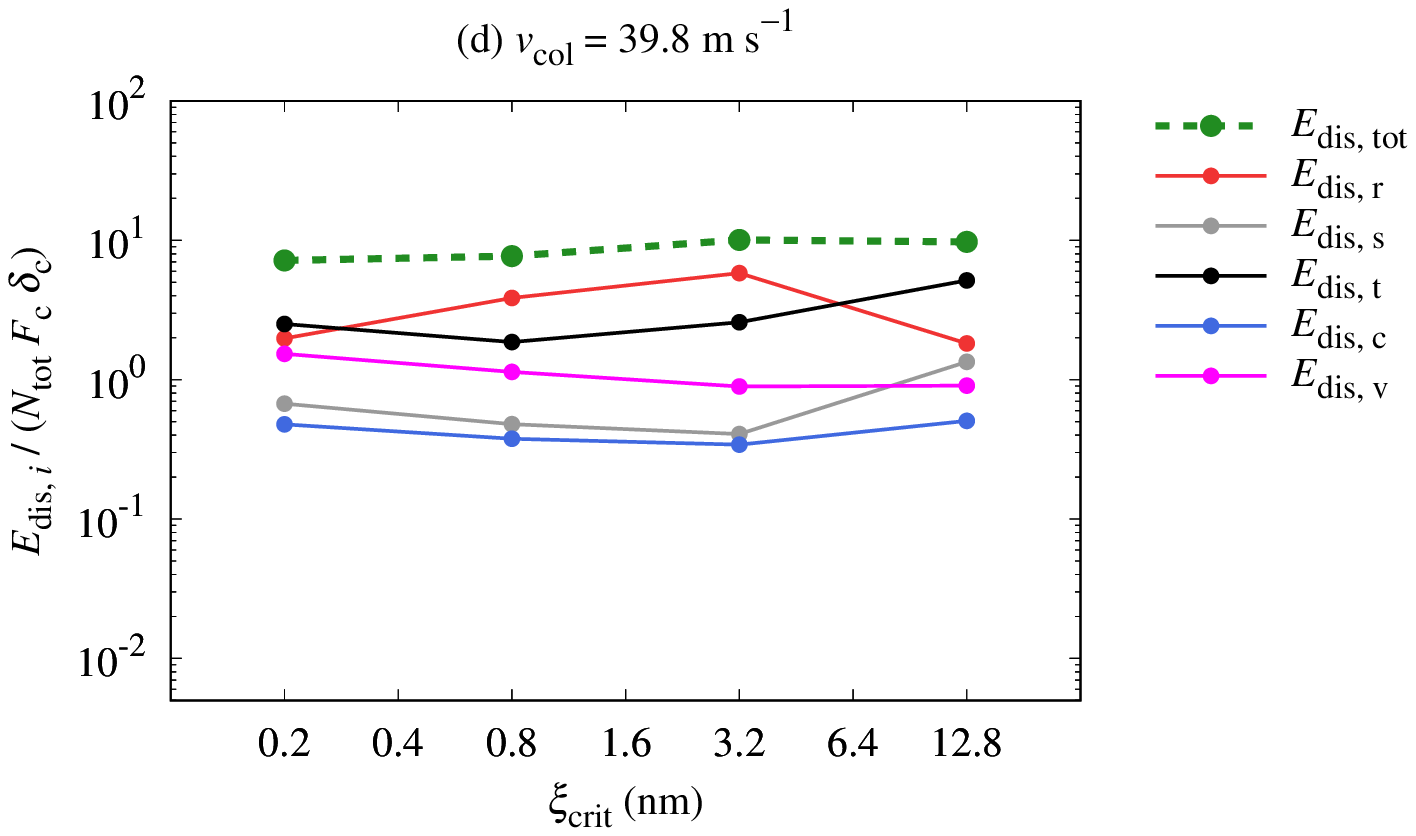}
\includegraphics[width=0.48\textwidth]{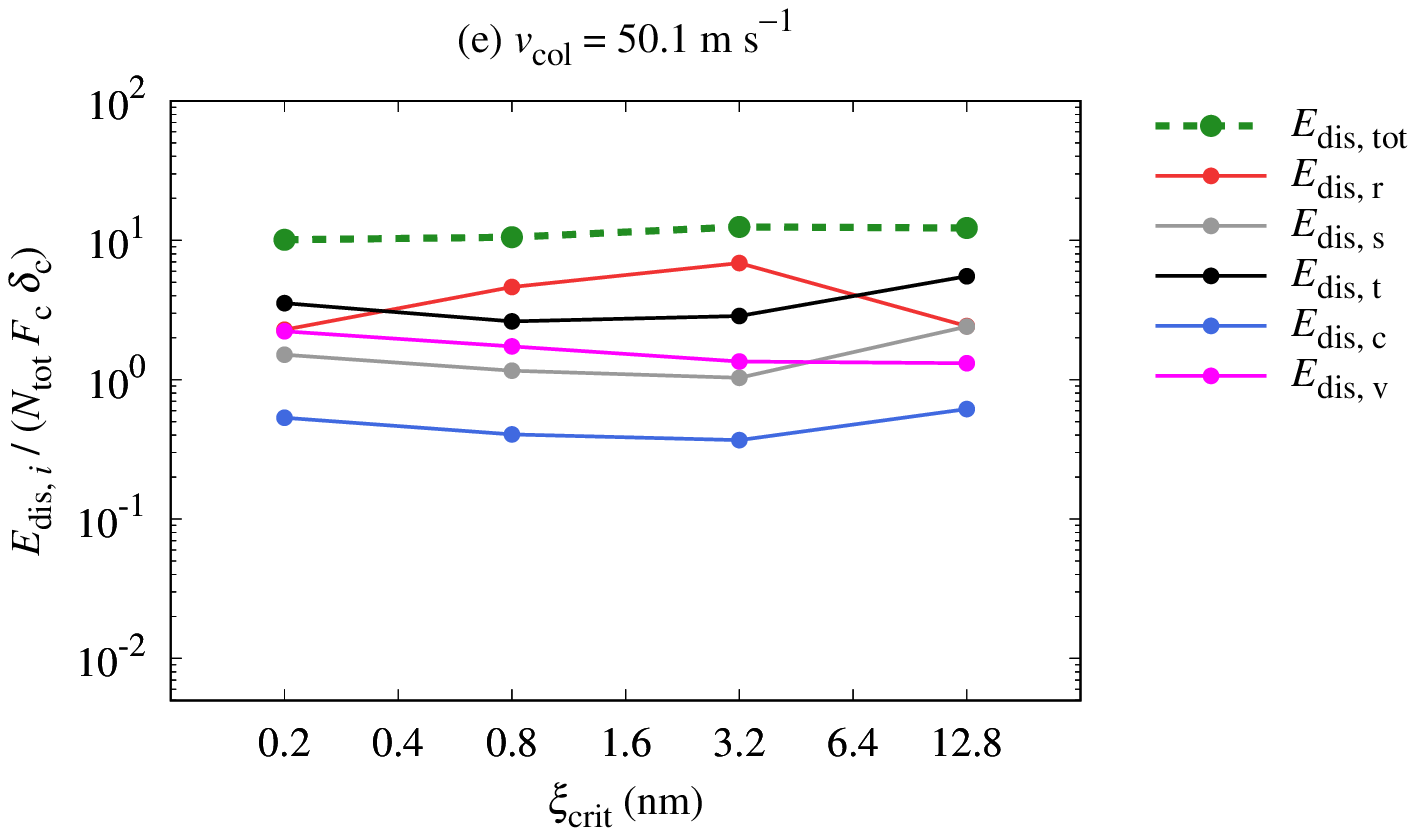}
\includegraphics[width=0.48\textwidth]{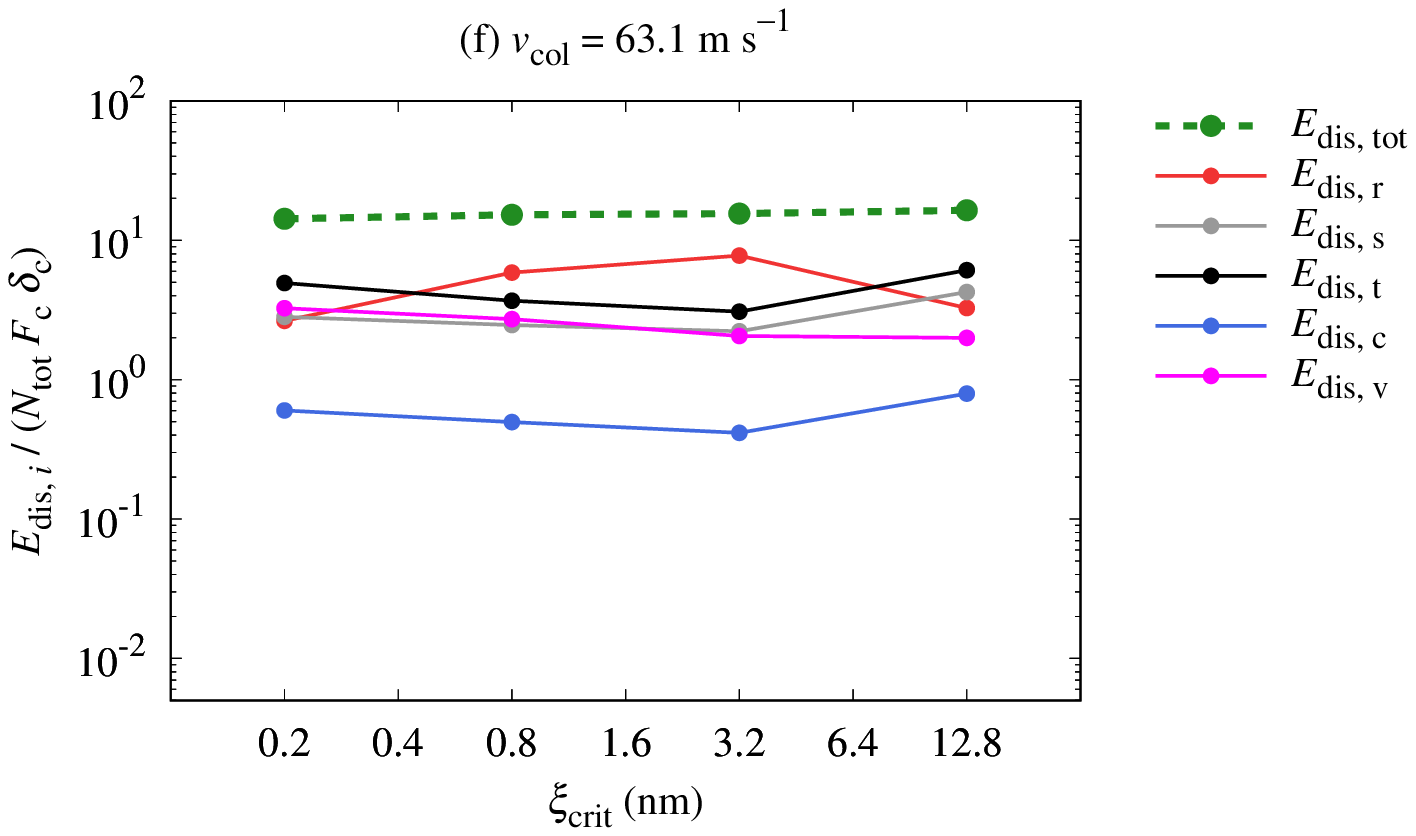}
\includegraphics[width=0.48\textwidth]{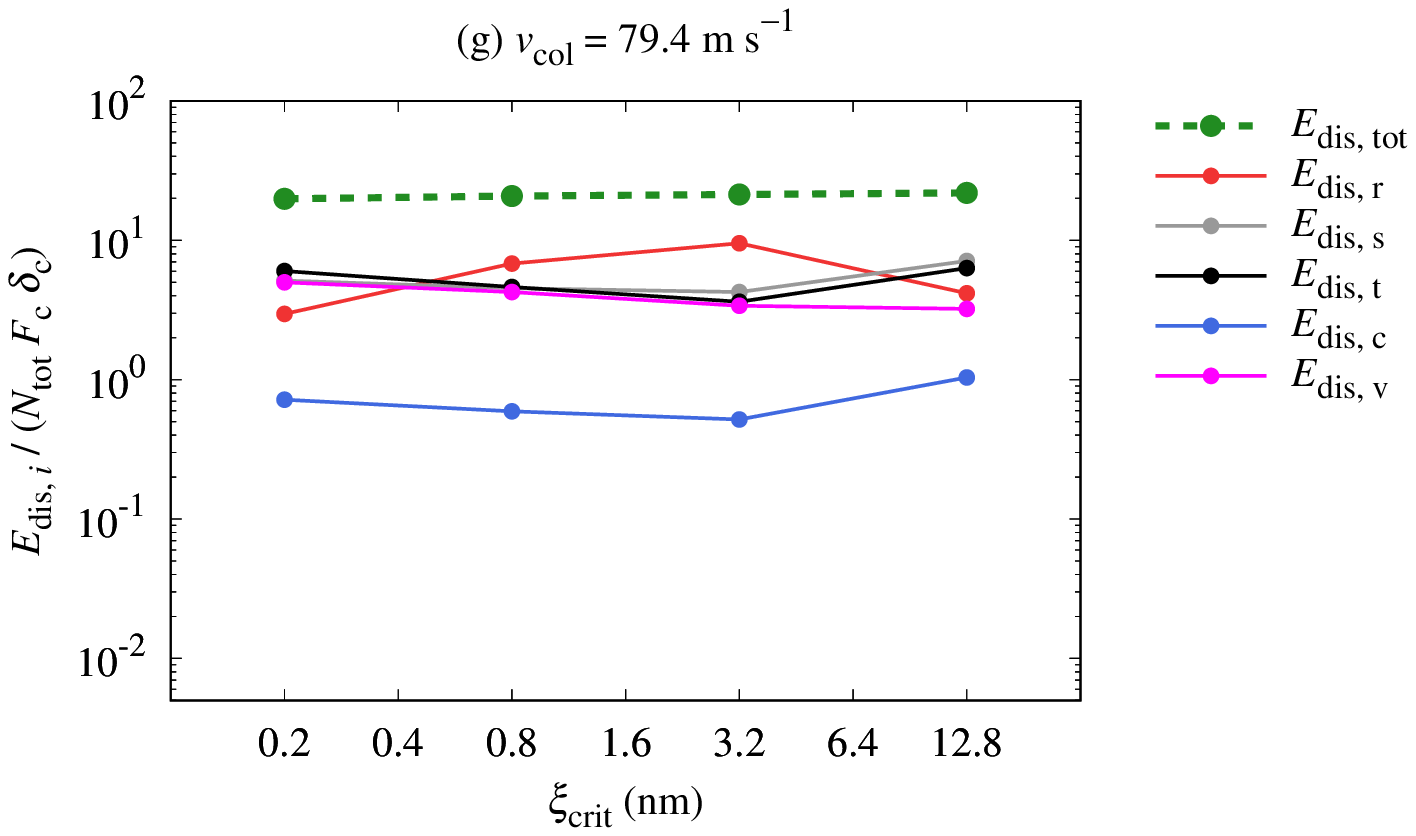}
\includegraphics[width=0.48\textwidth]{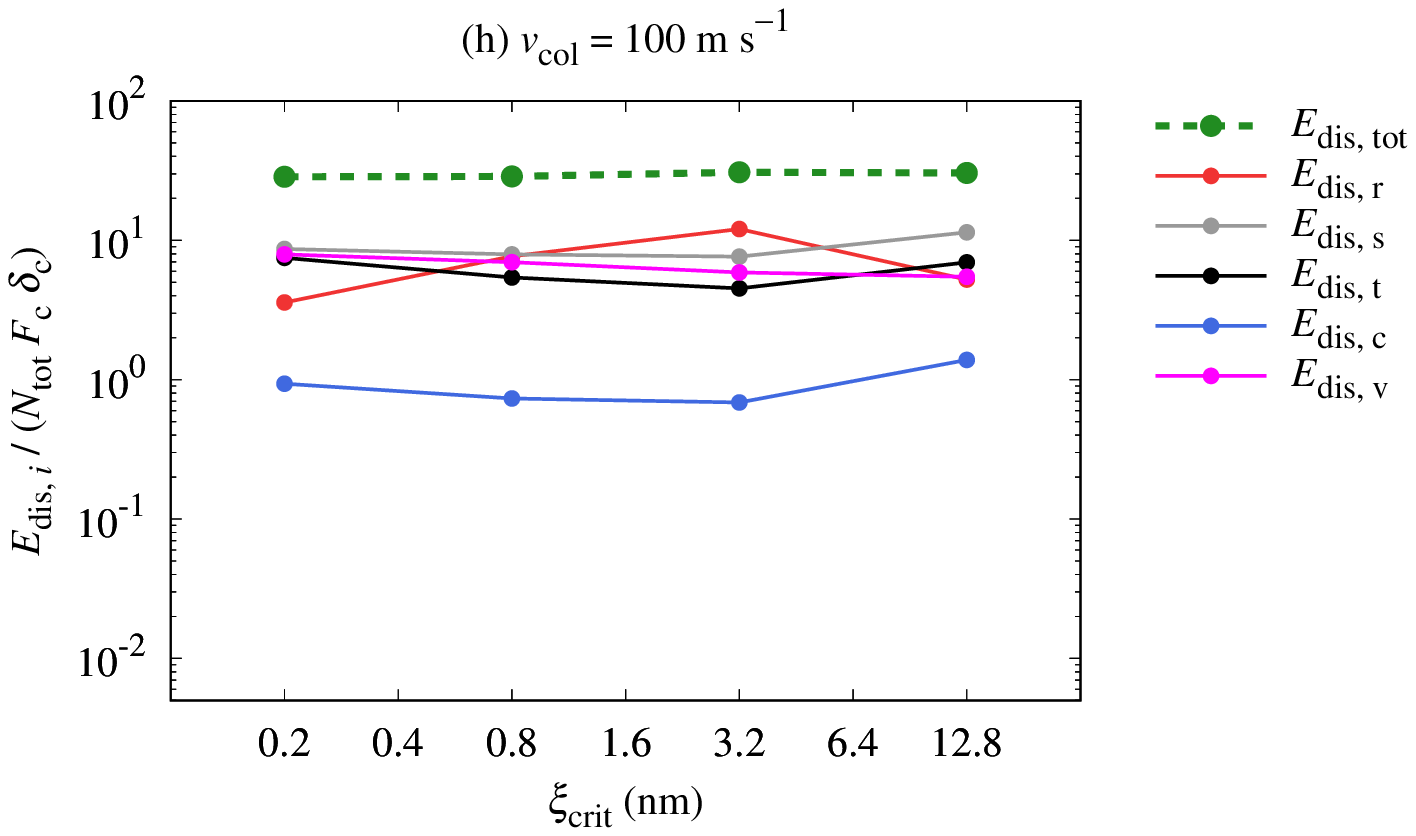}
\caption{
Same as Figure \ref{fig.edis} but with ${B_{\rm off}}^{2} = 6/12$.
}
\label{fig.edis6}
\end{figure*}

\begin{figure*}
\centering
\includegraphics[width=0.48\textwidth]{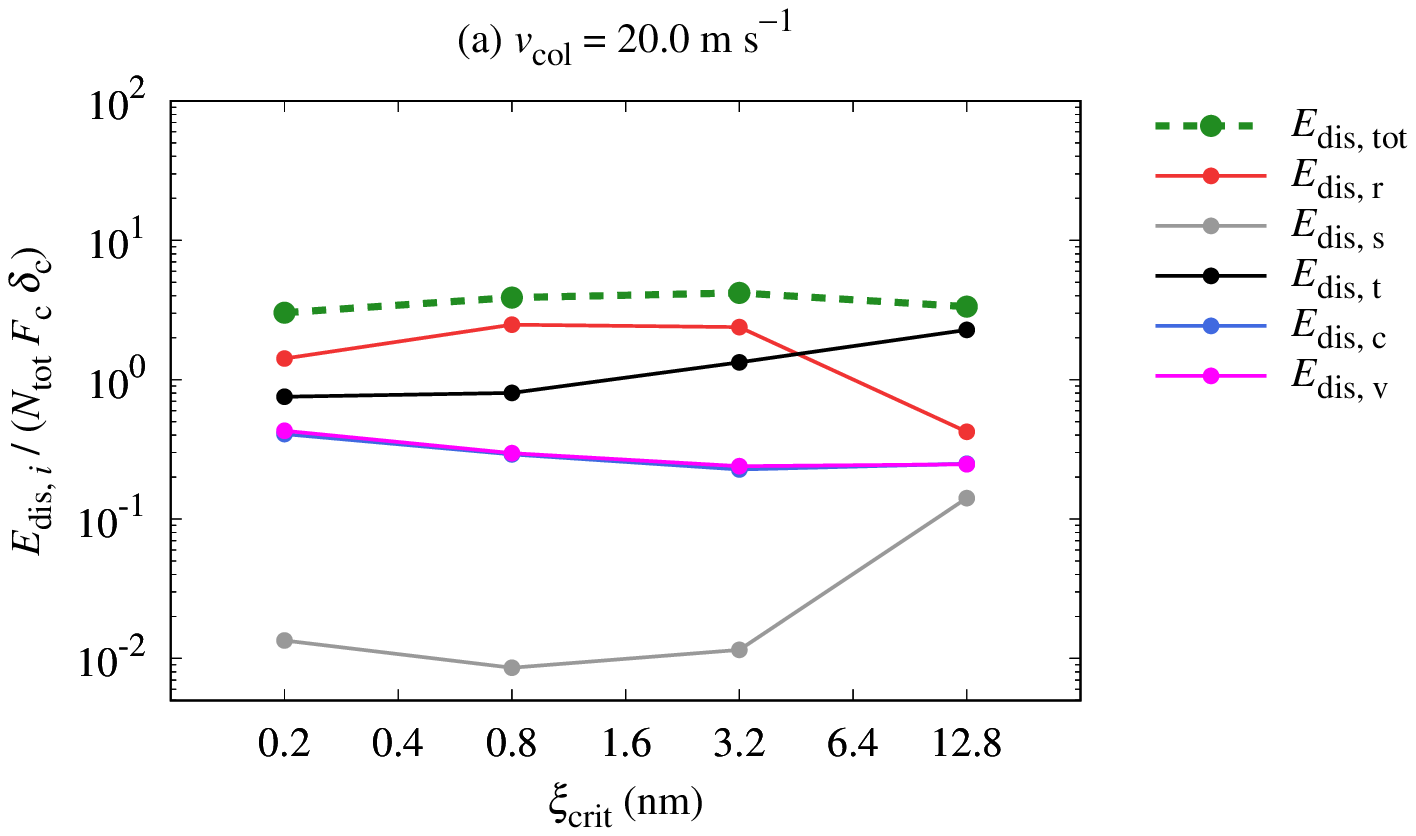}
\includegraphics[width=0.48\textwidth]{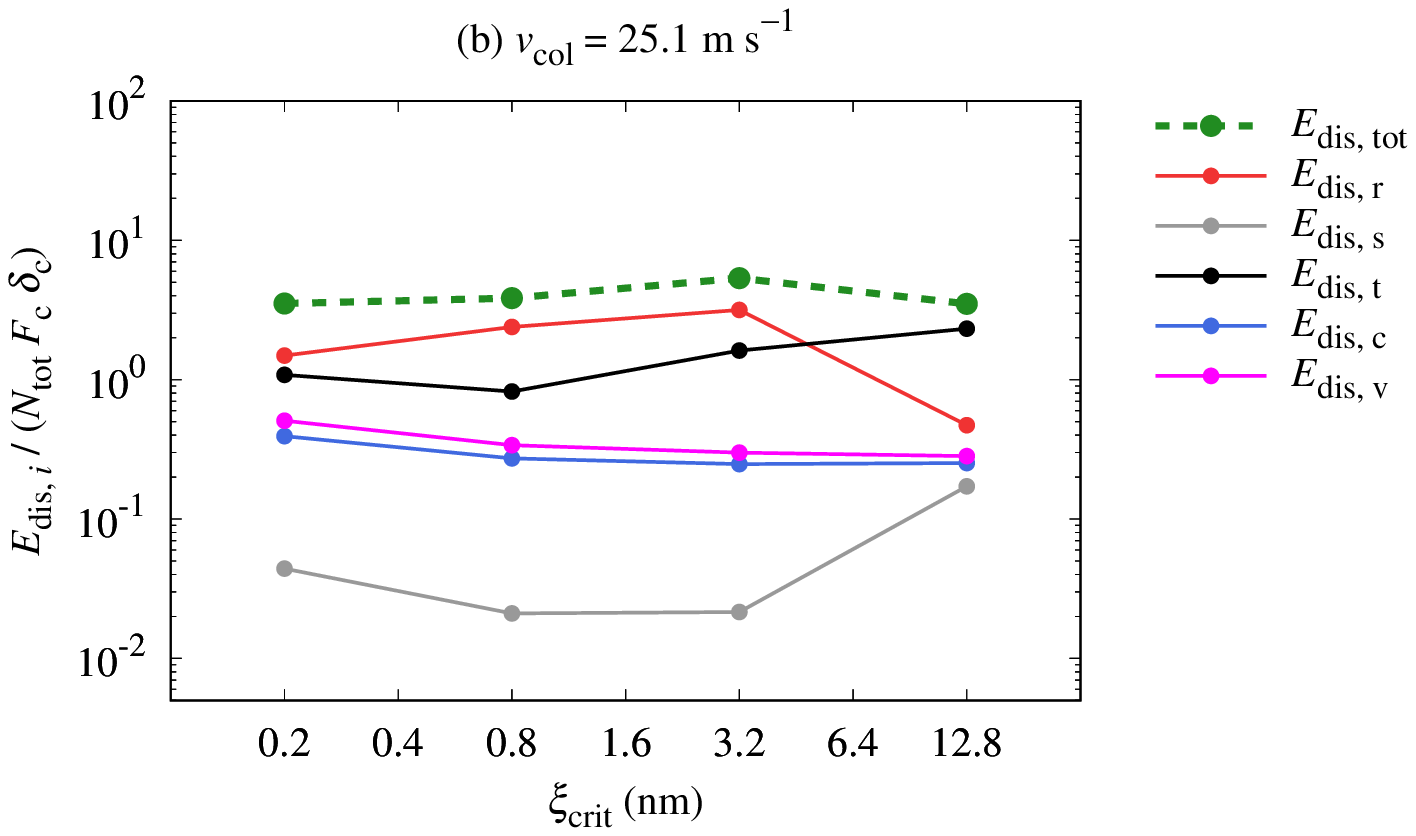}
\includegraphics[width=0.48\textwidth]{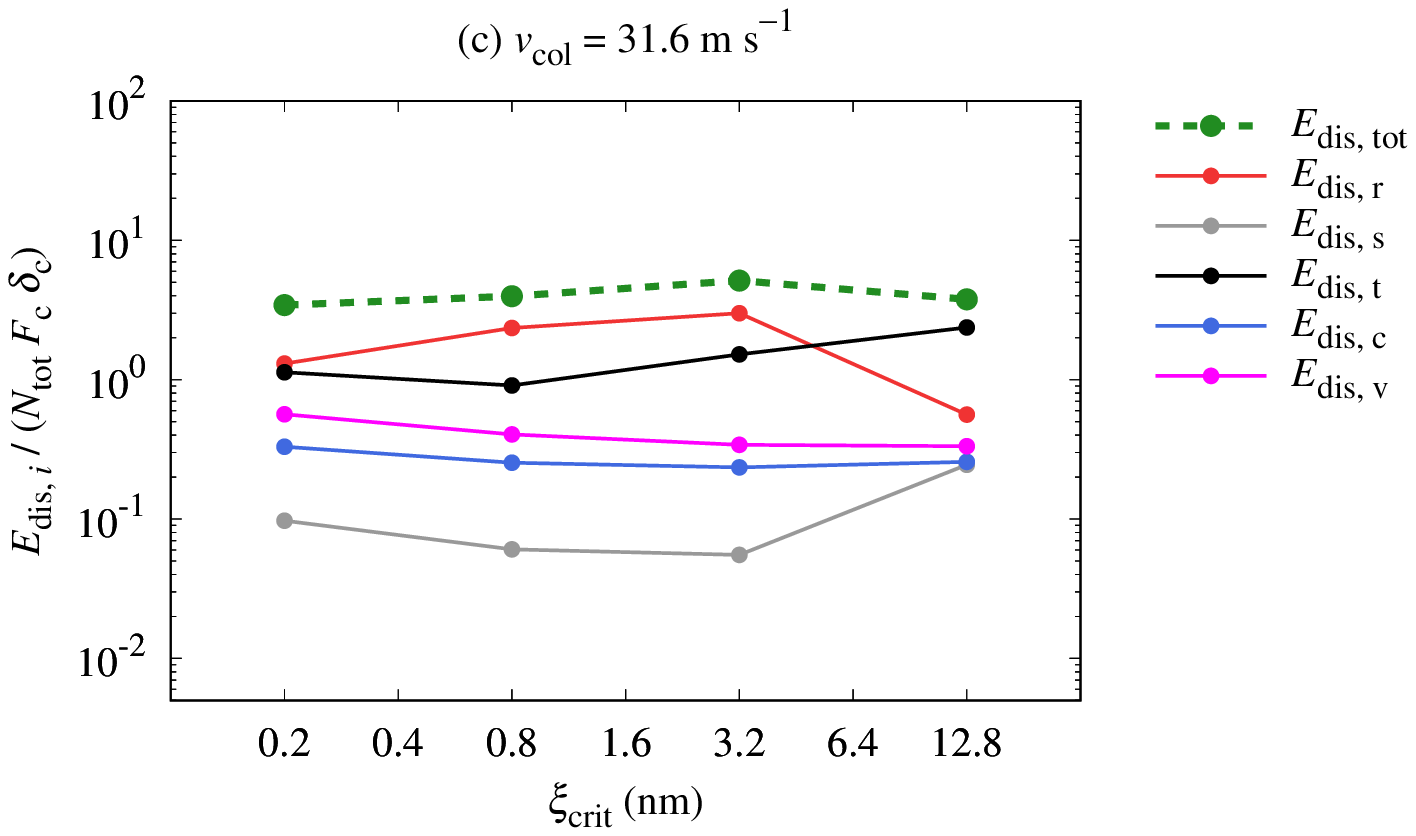}
\includegraphics[width=0.48\textwidth]{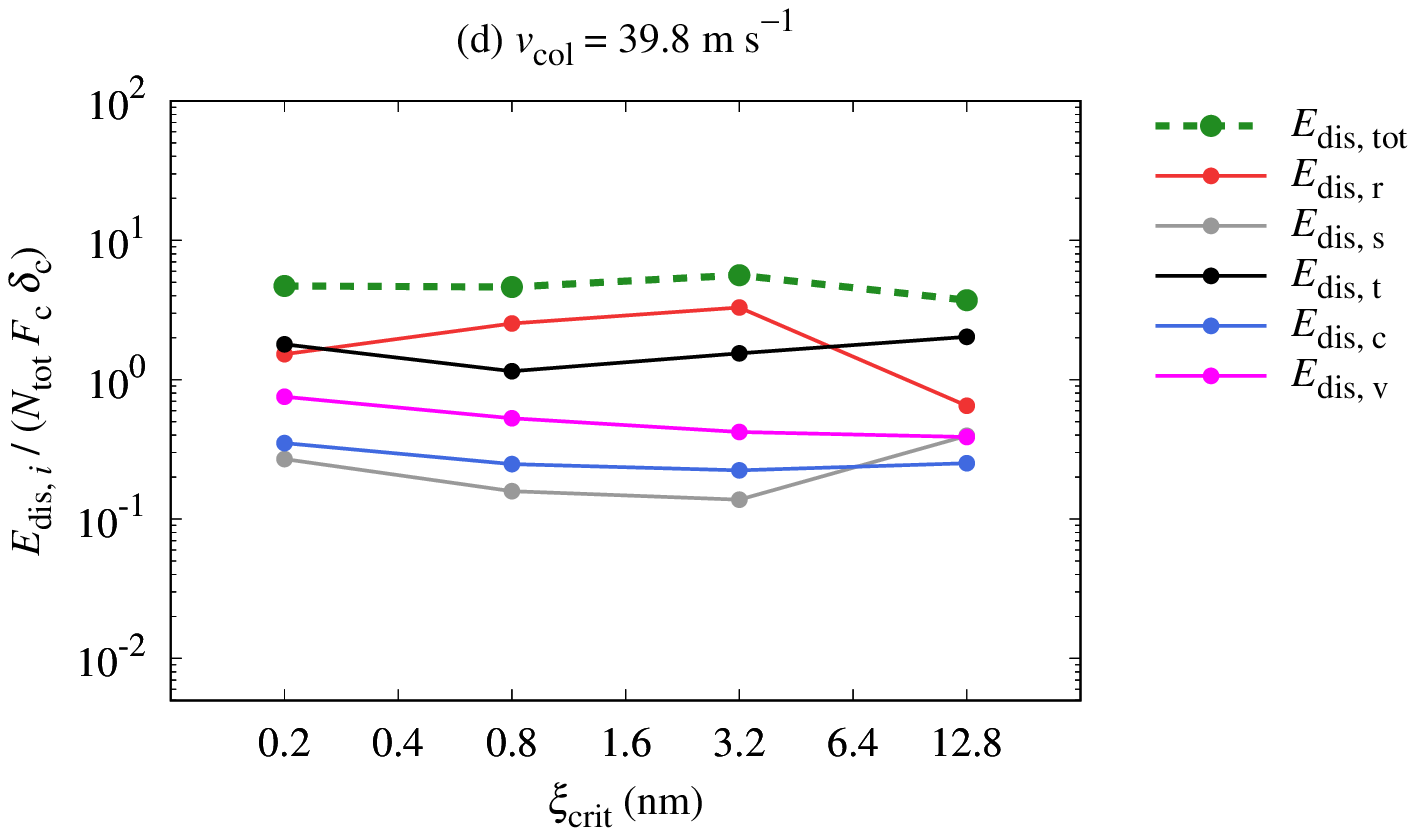}
\includegraphics[width=0.48\textwidth]{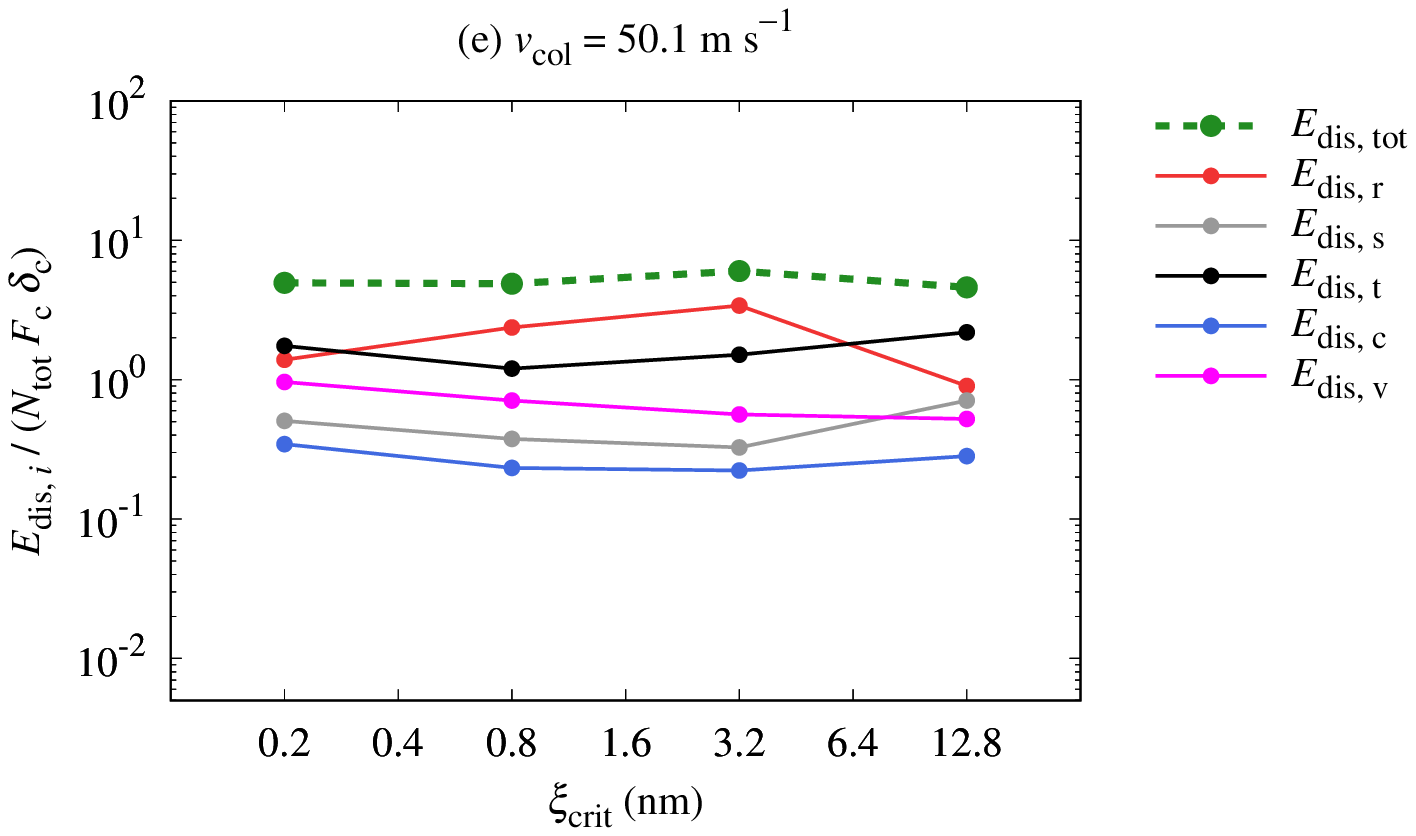}
\includegraphics[width=0.48\textwidth]{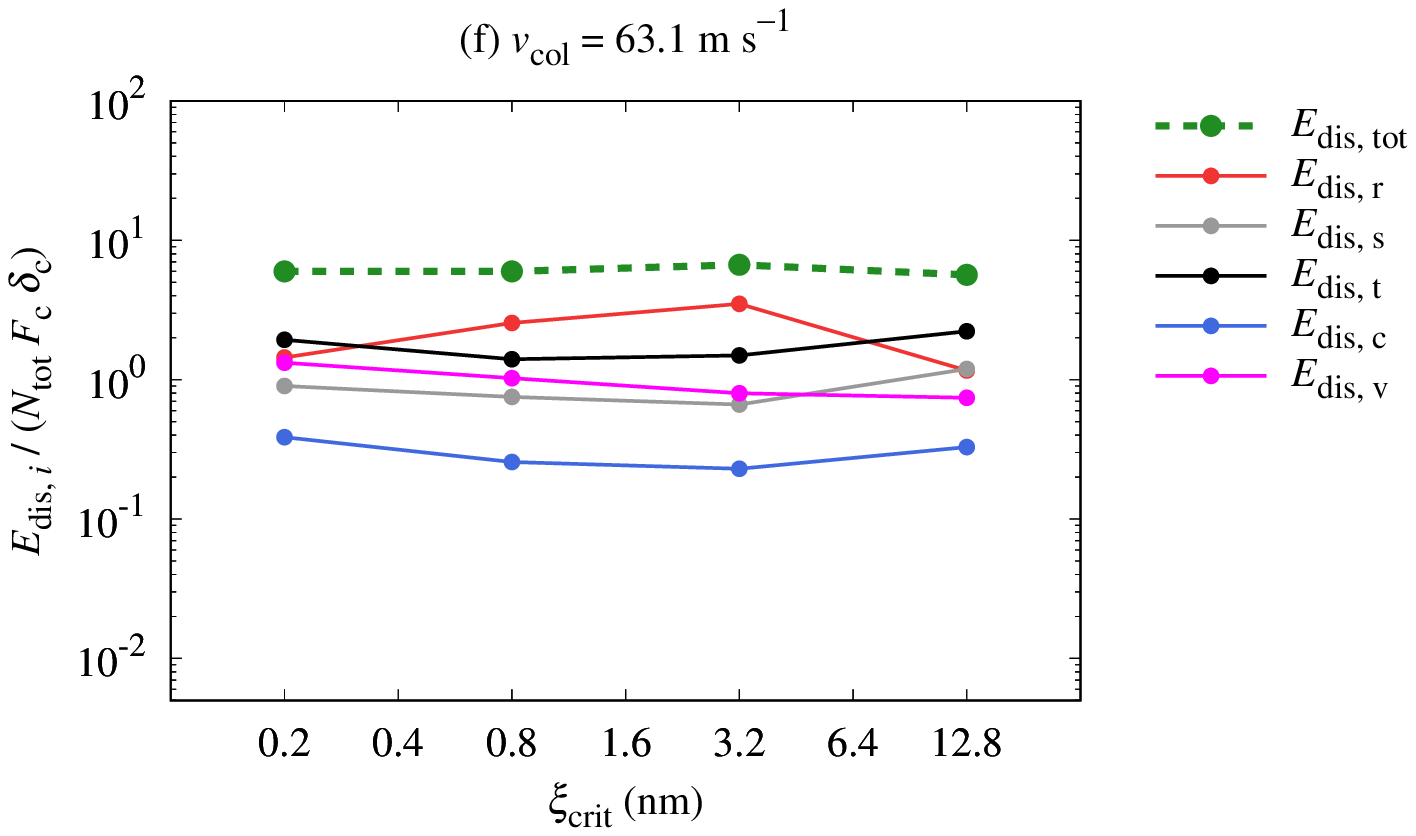}
\includegraphics[width=0.48\textwidth]{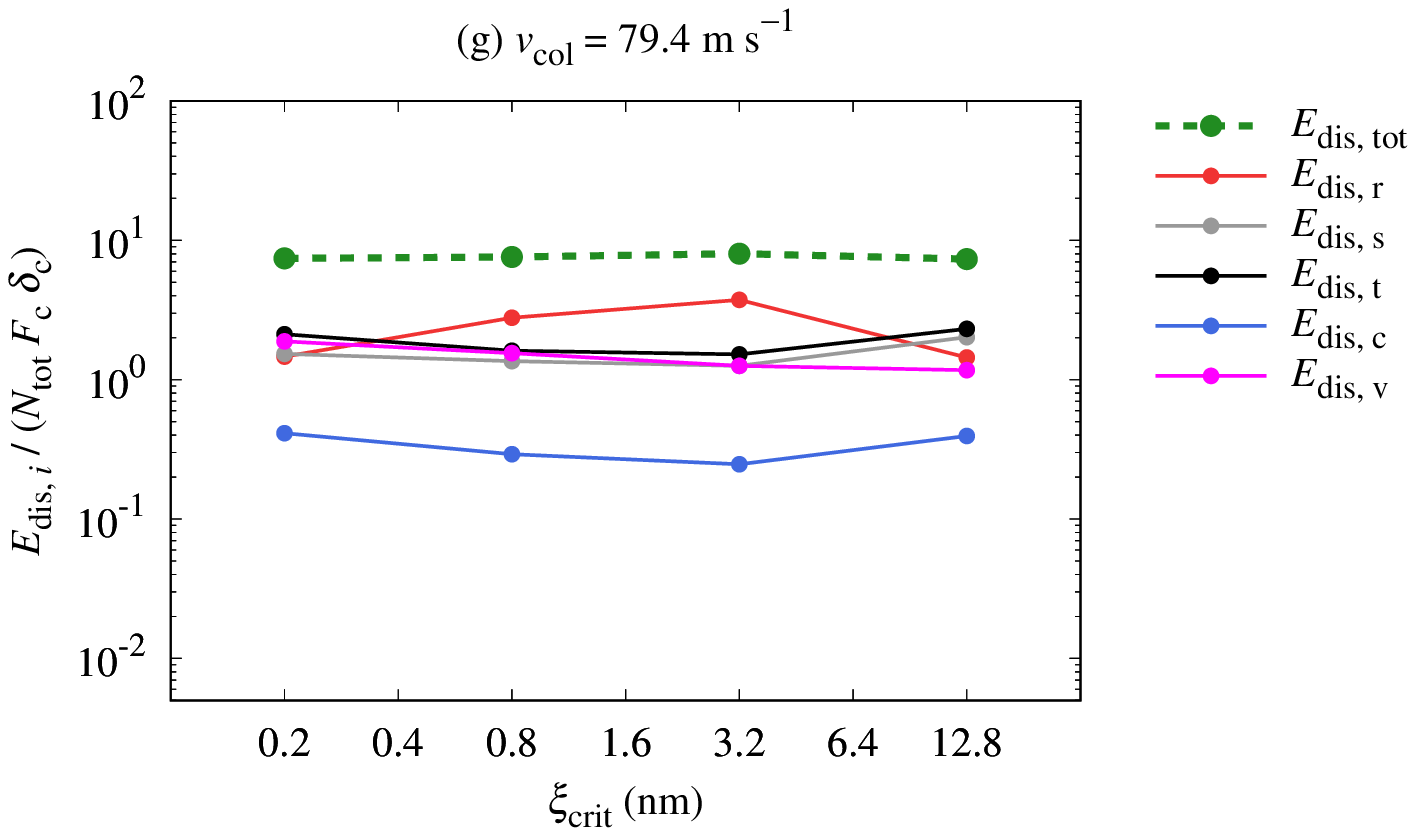}
\includegraphics[width=0.48\textwidth]{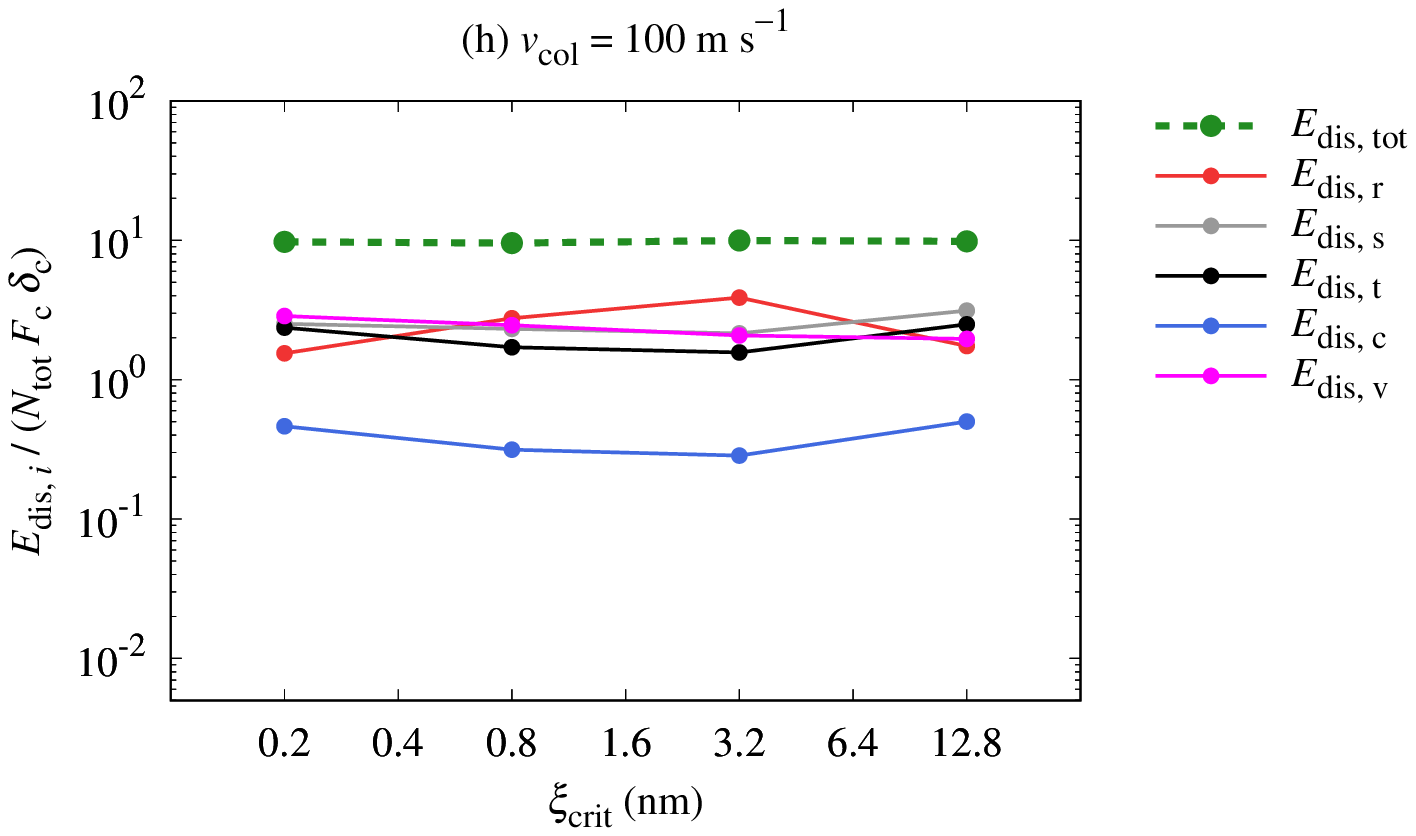}
\caption{
Same as Figure \ref{fig.edis} but with ${B_{\rm off}}^{2} = 9/12$.
}
\label{fig.edis9}
\end{figure*}

Figures \ref{fig.edis6} and \ref{fig.edis9} show the total energy dissipation due to particle interactions from the start to the end of the simulations.
We set ${B_{\rm off}}^{2} = 6/12$ and $9/12$ for Figures \ref{fig.edis6} and \ref{fig.edis9}, respectively.
The total energy dissipation due to all particle interactions, $E_{\rm dis, tot}$, is nearly independent of $\xi_{\rm crit}$ not only for ${B_{\rm off}}^{2} = 3/12$ but also for ${B_{\rm off}}^{2} = 6/12$ and $9/12$.

\begin{figure}
\centering
\includegraphics[width=\columnwidth]{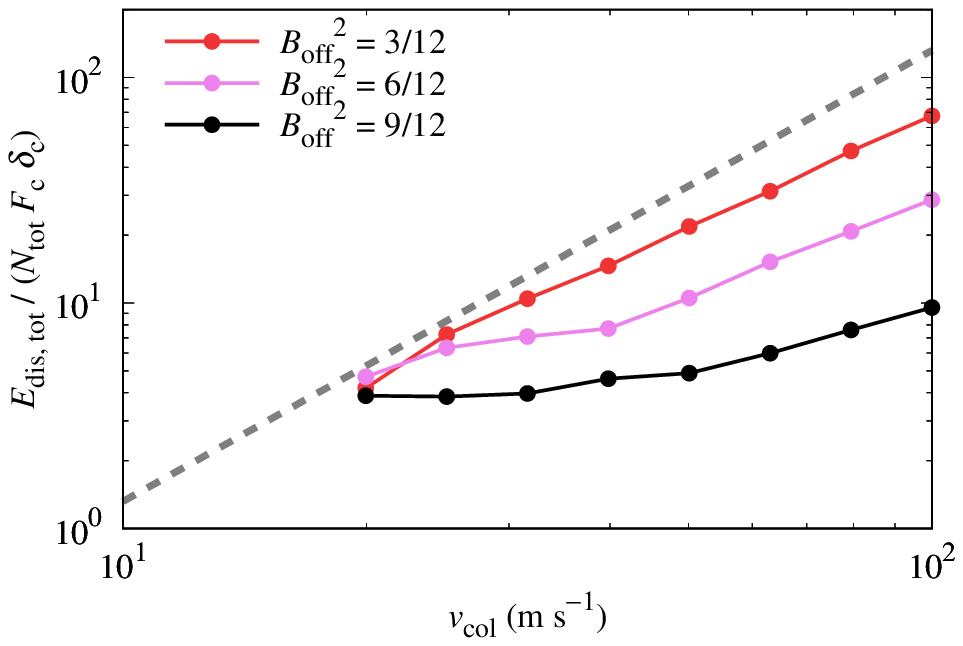}
\caption{
Dependence of $E_{\rm dis, tot}$ on ${B_{\rm off}}^{2}$ and $v_{\rm col}$.
Here we set $\xi_{\rm crit} = 0.8~\si{nm}$.
The dashed line is the initial kinetic energy, $N_{\rm tot} E_{\rm kin, 1}$, which is given by Equation (\ref{eq.E1}).
}
\label{fig.edis.vel}
\end{figure}

We note that $E_{\rm dis, tot}$ is smaller than the initial kinetic energy, $N_{\rm tot} E_{\rm kin, 1}$.
Figure \ref{fig.edis.vel} shows $E_{\rm dis, tot}$ as a function of ${B_{\rm off}}^{2}$ and $v_{\rm col}$.
Here we set $\xi_{\rm crit} = 0.8~\si{nm}$, although $E_{\rm dis, tot}$ is nearly independent of $\xi_{\rm crit}$.
The reason why $E_{\rm dis, tot}$ is smaller than $N_{\rm tot} E_{\rm kin, 1}$ is clear; fragments have kinetic energies after collisions as shown in Figure \ref{fig.snapshot}.

\bibliography{sample631}{}
\bibliographystyle{aasjournal}



\end{document}